\documentclass[prb,aps,amssymb,footinbib,showpacs]{revtex4-1}
\usepackage{amssymb}
\usepackage{amsmath}
\usepackage{times}
\usepackage{latexsym}
\usepackage{graphicx}
\usepackage{color}
\usepackage{bm}
\usepackage{subfigure}
\begin{document}
\title{ The Hofstadter Butterfly: Bridging Condensed Matter, Topology, and Number Theory }
\author{Indubala I Satija}
\affiliation{Department of Physics and Astronomy, George Mason University, Fairfax, VA 22030}
\date{\today}

\begin{abstract}

Celebrating its golden jubilee, the Hofstadter butterfly fractal emerges as a remarkable fusion of art and science. This iconic "X"-shaped fractal captivates physicists, mathematicians, and enthusiasts alike by elegantly illustrating the energy spectrum of electrons within a two-dimensional crystal lattice influenced by a magnetic field. Enriched with integers of topological origin that serve as quanta of Hall conductivity—this quantum fractal and its variations have become paradigm models for topological insulators, novel states of matter in 21st-century physics. This paper delves into the theoretical framework underlying butterfly fractality through the lenses of geometry and number theory. Within this poetic mathematics, we witness a rare form of quantum magic: Nature’s use of abstract fractals in crafting the butterfly graph itself. In its simplest form, the butterfly graph tessellates a two-dimensional plane with trapezoids and triangles, where the quanta of Hall conductivity are embedded in the integer-sloped diagonals of the trapezoids. The theoretical framework is succinctly expressed through unimodular matrices with integer coefficients, bringing to life abstract constructs such as the Farey tree, the Apollonian gaskets, and the Pythagorean triplet tree.

\end{abstract}

\maketitle

\section{Introduction}

\begin{center}
 {\it  ``All results of the profoundest mathematical investigation  must ultimately be \\ expressible in the simple form of properties of the integers. " --Leopold Kronecker}.
\end{center}

The enduring Power of Integers in unveiling  complexity as  prophesied by 
Leopold Kronecker
 finds a compelling illustration in the fascinating story of the Hofstadter butterfly and its connection to quantum phenomena.\\
 
 In 1975, as Benoît Mandelbrot was solidifying the concept of fractals \cite{M}, Douglas Hofstadter, then a graduate student, independently stumbled upon a remarkable quantum fractal \cite{Hof}. This fractal described the energy spectrum of electrons in a two-dimensional crystal subjected to a magnetic field, characterized by its distinctive "X" shape, reminiscent of a butterfly nested infinitely within itself, earning it the name "Hofstadter butterfly." The elegant complexity and inherent order of the Hofstadter butterfly captivated Gregory Wannier \cite{W} and Francisco Claro \cite{CW}. In 1978, they revisited the problem, providing a simplified representation of the spectrum, now known as the Wannier diagram, and crucially integrated integers into the mathematical framework of the butterfly fractal. These integers came to life in $1982$ as they were identified with the quanta  of conductivity\cite{Streda, TKKN} in the quantum Hall effect discovered in 1980 \cite{QHEK}. Furthermore, seminal work by Thouless and collaborators \cite{TKKN} demonstrated that this quantization is rooted in topology. This groundbreaking discovery was the first example of topological insulators \cite{Spinbook}, an exotic class of matter that is revolutionizing condensed matter physics in this century. The profound implications of this discovery led to David Thouless being awarded the Nobel Prize in Physics in 2016.\\
  
This paper embarks on a journey to reveal the power of integers  in describing the intricate Hofstadter butterfly fractal. Before delving into the technical specifics, we provide a summary to showcase the inherent richness and beauty of this problem, which explores the behavior of electrons in a two-dimensional crystal under the influence of a magnetic field. The essence of this story is captured in two sets of figures, followed by brief description. Figures  offer pictorial highlights of the butterfly graph, showcasing its role as a bridge between condensed matter physics, topology, and number theory, thereby revealing its multifaceted nature as described below. These images serve as a guide to the theoretical framework, uncovering the simplicity hidden within the complexity of this quantum object.

  \begin{figure}[htbp] 
  \includegraphics[width = .85 \linewidth,height=.53 \linewidth]{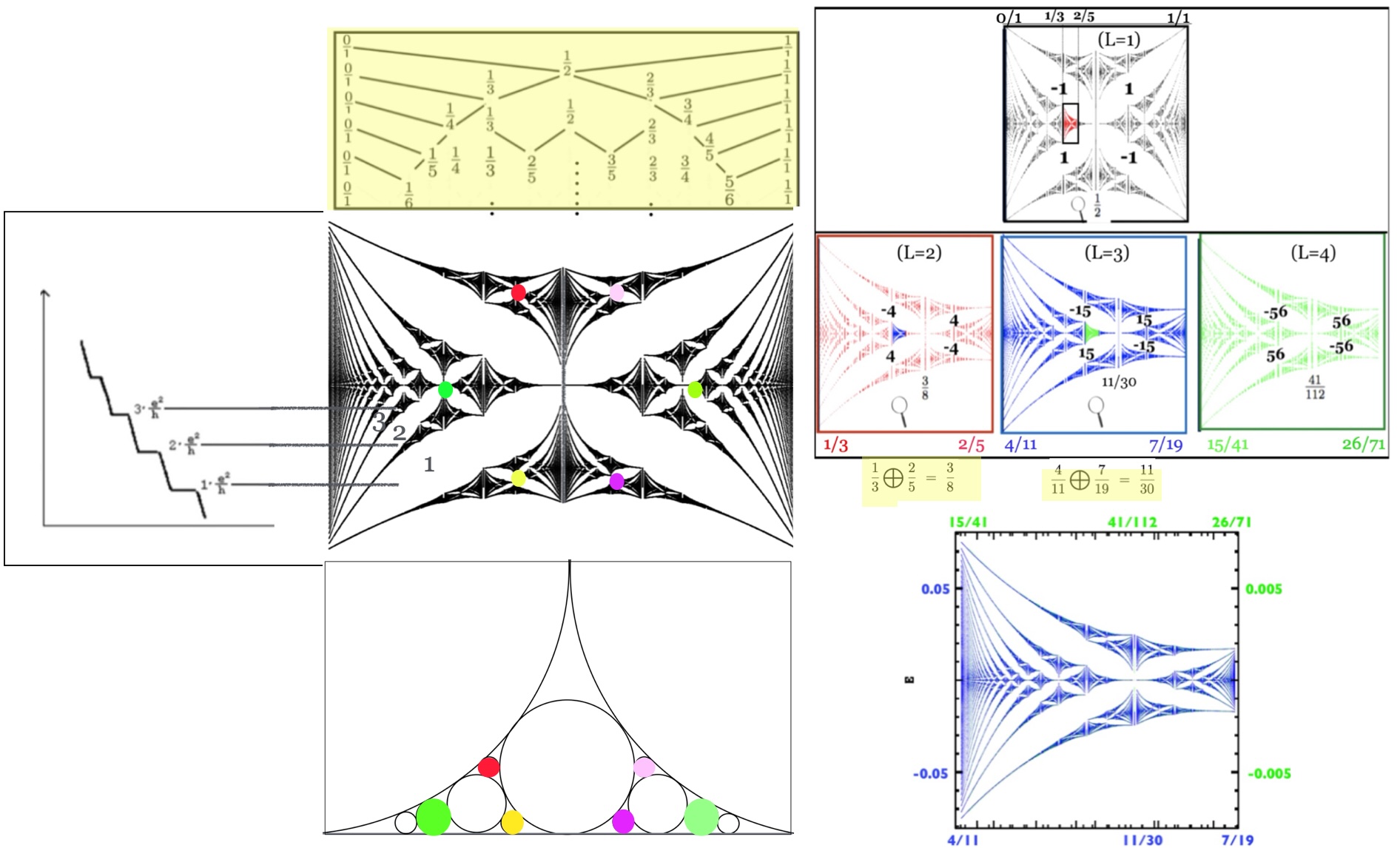} 
 \caption{ The butterfly fractal (middle vertical panel) displays wings (gaps), some  labeled with integers that correspond to the plateaus of Hall conductivity (left). The Farey tree, highlighted in yellow, serves as the structural backbone of this fractal, connecting it to the Apollonian gasket shown at the bottom. Right panel illustrates self-similar
 character of the butterfly and highlights  ( in yellow ) the fact that  each butterfly adhers to the "Farey sum rule" (Equation (\ref{FR})). Six color-coded circles within the butterfly and Apollonian indicate the recursive structure of these fractals, which is further detailed in the next figure. }
  \label{b1}
\end{figure}

 \begin{figure}[htbp] 
   \includegraphics[width = .75 \linewidth,height=.45 \linewidth]{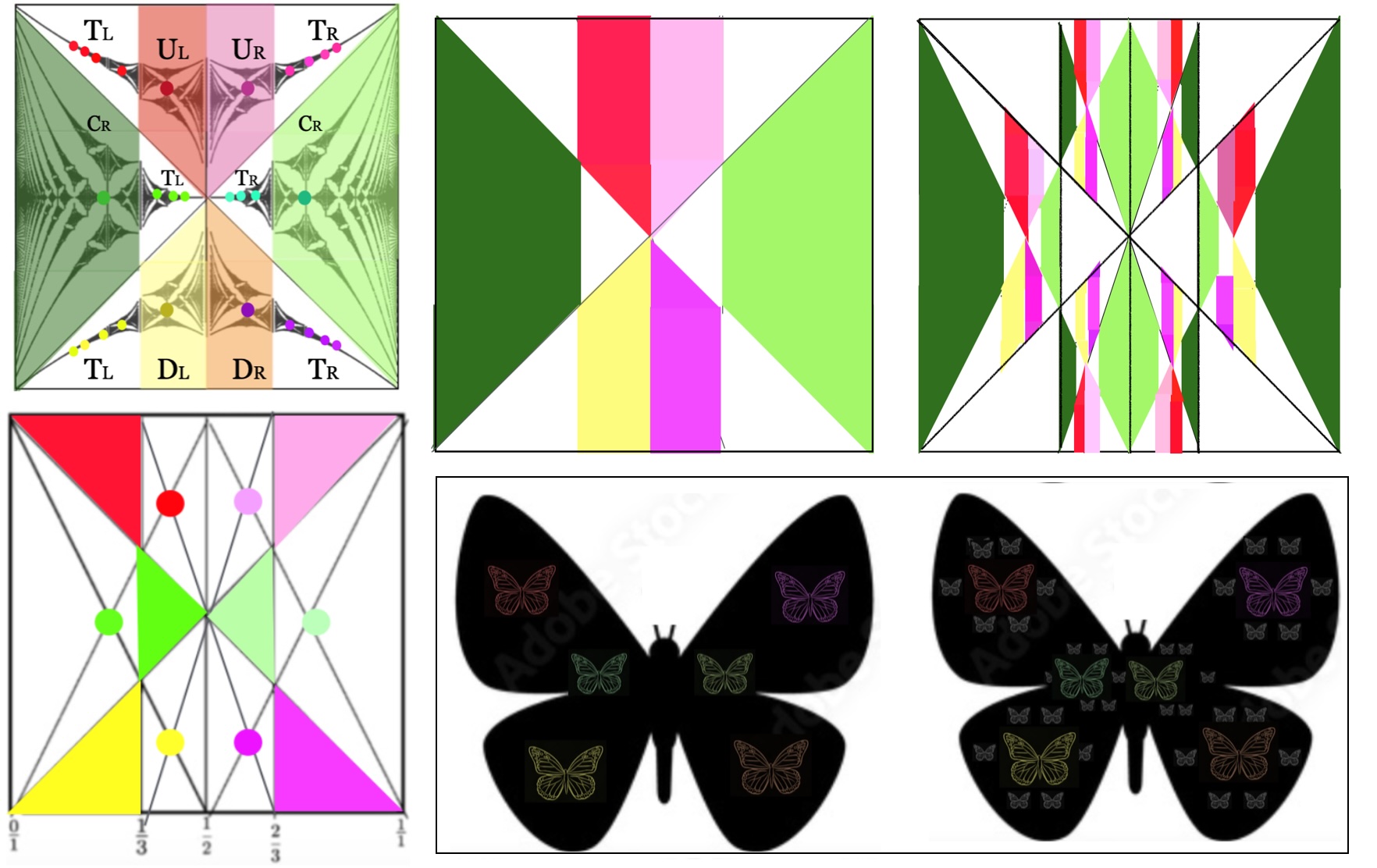} 
 \caption{ The figure serves as a guide to a simplified representation of the butterfly graph, envisioned as a 2D tessellation of trapezoids and triangles. The iterative scheme underlying this fractal involves a parent butterfly generating six baby butterflies, labeled as $(U_L, U_R, D_L, D_R, C_L, C_R)$,  each accompanied by a tail —an infinite chain of butterflies.  Chains
 ( color coded with the butterfly )  are denoted as $(T_L,T_R)$ as only two generators are required to generate them.  In essence, a "butterfly with a tail" (where butterflies are depicted as trapezoids and tails as triangles) forms the building blocks of the butterfly graph. The right panels illustrate two generations of the recursions, depicted both as a hierarchical lattice (upper panel) and a schematic (lower panel).}
  \label{b2}
\end{figure} 

Below is a brief synopsis of the narrative that leads us through the theoretical framework, revealing the integer wonderland at the heart of the butterfly fractal. It provides insights into Nature's use of abstract fractals to illuminate the enigmatic microscopic world of electrons within a crystal, responding to competing length scales- the periodicity of the crystal and the magnetic length related to cyclotron orbits in the magnetic field.

\begin{itemize}
\item {\bf Relation to Quantum Hall States}: Figure (\ref{b1}) emphasizes a crucial aspect of the butterfly graph—its link to the exotic phenomenon of Hall conductivity quantization discovered in 1980. The integer quanta intricately weave through the butterfly graph, forming the building blocks of its recursive patterns. This highlights the importance of fractals as not merely abstract mathematical constructs, but as essential components of quantum physics.
\item {\bf Relation to Farey Tree}: The backbone of the butterfly fractal is an abstract mathematical construct—the Farey tree \cite{book}, discovered nearly one century before the butterfly graph by Adolf Hurwitz in 1894. As depicted in the figure, the Farey tree generates all primitive rationals between 0 and 1. This tree is brought to life in the butterfly graph as all butterflies adhere to the "Farey sum formula":
\begin{equation}
 \frac{p_c}{q_c} = \frac{p_L+p_R}{q_L+q_R} 
 \label{FR}
 \end{equation}

where the three fractions in the triplet $\Big[\frac{p_L}{q_L}, \frac{p_c}{q_c}, \frac{p_R}{q_R}\Big]$ represent the magnetic flux values at the left, center, and right boundaries of each butterfly in the butterfly graph. This equation is a byproduct of the renormalization group (RG) \cite{SW} and suggests an intertwining of number theory with the quantum mechanics of Bloch electrons in a magnetic field. The middle panel, displaying self-similar butterfly images, shows that the "Farey sum rule" is embedded at all scales in the butterfly graph.
\item {\bf Relation to Apollonian Gasket}: The Apollonian gasket is a close packing of four circles \cite{IAP} with integer curvatures (reciprocal of the radii), where each circle is tangent to three others. It is indeed amazing that the recursive structure of this beautiful abstract set, discovered in 200 BC, mirrors the butterfly recursions.
\item{ \bf 2D Tessellation of Trapezoids and Triangles}: The skeleton form of the butterfly graph is described as a tessellation or tiling of trapezoids and triangles in two dimensions \cite{SAT25}. A unique aspect of this tessellation is that the diagonals of these trapezoids have integer slopes, reincarnations of the quanta of Hall conductivity. This leads to an eightfold mathematical framework for constructing the fractal: The butterfly graph is dissected into sextuplets of butterflies, labeled as $(U_L, U_R, D_L, D_R, C_L, C_R)$. These labels denote six generators that are unimodular matrices with integer coefficients. Furthermore, each butterfly in the graph has an attached "butterfly tail," constructed using two additional generators, forming an eightfold prescription necessary to build the entire butterfly fractal.
\end{itemize}

\section{ Electrons in a magnetic field: from Landau levels to  Butterfly Fractal\cite{review}}

The study of the quantum behavior of electrons in a magnetic field has a rich and extensive history, marked by contributions from pioneering figures such as Lev Landau (1930), Rudolf Peierls (1933), Joaquin Luttinger (1951), Lars Onsager (1952), Walter Kohn (1962), and many others\cite{book}. The hierarchical nature of this system’s spectrum became apparent in the 1960s\cite{azbel, others}, drawing mathematicians into the field. The topic gained significant attention in 1981 when Mark Kac famously offered a "ten martini" prize to anyone who could prove that the butterfly spectrum is a Cantor set of zero measure\cite{ten}.\\

With the advent of quantum mechanics, the theoretical exploration of electrons moving in a two-dimensional crystal subjected to a transverse magnetic field achieved three major turning points: Bloch's theorem (1928), Landau levels (1930), and Peierls' substitution (1933)\cite{review, bookadd}. In 1955, these foundational works culminated in Harper's equation—the model system that generates the butterfly spectrum. We begin by examining Landau's pioneering work, which focused on investigating diamagnetism in electrons.

\subsection{  Landau Levels}

The Hamiltonian describing an electron of mass $m$ and charge $q$ moving  x-y plane in the presence of a uniform transverse magnetic field  $B\hat{z}$ is given by\cite{review}:
\begin{equation}
H = \frac{(p_x - q A_x )^2}{2m} +  \frac{(p_y - q A_y )^2}{2m} 
\label{qL}
\end{equation}
where $\vec{B} = \nabla \times \vec{A}$. 
Classically, the system describes electrons moving in circular orbits called the cyclotron orbits with frequency,
\begin{equation}
w_c  =   \frac{qB}{m}   
\end{equation}

This problem presents a fascinating aspect: regardless of the size of the circle, electrons take the same amount of time to complete a revolution. In a gas of non-interacting electrons moving in two dimensions (2D) under the influence of a transverse magnetic field, the electrons behave like disciplined terra cottas, moving in synchronized circles, each executing one revolution in the same duration.\\

As a quantum Hamiltonian,  Eq. (\ref{qL}) is equivalent to a one-dimensional harmonic oscillator  as
the two non-commuting operators  $( (p_x - q A_x ), (p_y - q A_y ))$ satisfy the commutation relation:
\begin{equation}
 [( (p_x - q A_x )/\hbar, (p_y - q A_y )/\hbar)]=  \frac{ i q B}{\hbar}.
 \label{cr}
 \end{equation}
 
Therefore, we can identify these two operators as canonical variables $(\hat{x}, \hat{p})$  satisfying  $ [ \hat{x}, \hat{p}  ] = \frac{ i q B}{\hbar}$  with effective Planck constant equal to $\frac{ i q B}{\hbar} $.
Therefore, Landau Hamiltonian can be written as:
 \begin{equation}
H_L =  \frac{\hbar^2}{2m} ( \hat{p}^2 + \hat{x}^2 ), \quad [ \hat{x}, \hat{p} ] = \frac{ i q B}{\hbar}
\label{lh}
\end{equation}

In other words, the two-dimensional problem of an electron with transverse magnetic field is mapped to one-dimensional harmonic oscillator 
 with  effective Planck's constant $qB/\hbar $. The energy levels of the system are given by:
\begin{equation}
E_n = \hbar \omega_c ( n+ \frac{1}{2} ). 
\end{equation}

These energy levels are known as Landau levels.  
We note that  a new length scale emerges in the quantum problem that characterizes exponentially decaying
 wave functions of the electron. This ``magnetic length", denoted as $l_B$  is given by:
   \begin{equation}
  l^2_B =  \left(\frac{\hbar}{q} \right)\frac{1}{B}
  \end{equation}
  
  As we will explore next, when electrons are subjected to a periodic potential while traversing a crystal, the magnetic length competes with the crystal's periodicity. This competition lies at the core of the fractal nature of the butterfly spectrum.
  
\subsection{ Electrons in a 2D Periodic Potential- Harper Equation }

In the presence of a 2D crystal, the  Hamilton is given by,
\begin{equation}
H =  \frac{1}{2m}[ ( (p_x - q A_x )^2+(p_y - q A_y )^2] + V(x, y) ],
\end{equation}
where $V(x, y)$ is a periodic function of $\vec{r}=(x,y)$. If the magnetic field is strong, the periodic potential can be treated as a perturbation on Landau Hamiltonian.
Alternatively, if the magnetic field is weak, we can start with an electron in a periodic potential and treat the magnetic field as a perturbation.\\

Theoretical study of electrons in a periodic potential originated in $1928$  foundation work 
 by Swiss physicist Felix Bloch. Known as Bloch's theorem, it  states that the wave function of a free electron in a periodic potential is given by,
 \begin{equation}
 \psi (\vec{r} ) = e^{ i \vec{k} \cdot \vec{r} } u_k (\vec{r}),
 \end{equation}
 where $u_k(\vec{r})$ shares the periodicity of the lattice. Simplest description is to consider  "tight binding approximation"  where an electron on a lattice can hop only
 to its nearest neighbors.  In this case, the Bloch's theorem gives one band model of the energy level. For a rectangular lattice with $ (t_x, t_y)$ being
  the hopping amplitudes along the x and y direction, energy levels of the electron are,
 \begin{equation}
 E(k_x,k_y) = [  t_x \cos (k_x a ) + t_y \cos(k_y a) ],
 \label{band1}
 \end{equation}
 
To introduce magnetic field into this tight binding model, one uses
 ``Peierls's substitution" - an important advancement introduced by Peierls in $1933$\cite{PBB}  where crystal momentum $\hbar \vec{k}$ is replaced by $\hbar \vec{k} - q \vec{A}$. In $1955$, Philip  Harper used Peierls substitution for the one band model to derive the so called Harper equation\cite{harper} as described below.\\

With Peierls substitution: $\hbar \vec{k} \rightarrow ( \vec{p} - q \vec{A}) $,  and using Landau gauge $\vec{A} = ( 0, Bx)$,   Eq. (\ref{band1})  is transformed to a tight binding Hamiltonian,
\begin{equation}
H_{\rm tbm} =  t_x [ e^{ \frac{ ip_x a}{\hbar} }  + e^{ \frac{-ip_x a} {\hbar} }] +  t_y [e^{ \frac{ (ip_y - q Bx) a}{\hbar} }  + e^{ \frac{-(ip_y-qxB )  a} {\hbar} } ], 
\end{equation}
and the eigenvalue equation is,
\begin{equation}
H_{\rm tbm} \psi(x,y) = E \psi(x,y)
\end{equation}
The $(p_x, p_y)$ are translation operators and therefore,
\begin{equation}
e^{ \frac{ ip_x a}{\hbar} } \psi(x,y) = \psi(x+a,y) ,\,\,\,\ e^{ \frac{ ip_y a}{\hbar} } \psi(x,y) = \psi(x,y+a)
\label{tr}
\end{equation}
Therefore, the eigenvalue equation can be written as:
\begin{equation}
\psi(x+a,y) + \psi(x-a,y)+ \lambda [e^{- 2 \pi i \phi} \psi(x,y+a) + e^{ 2 \pi  i \phi} \psi(y+a)]  = E \psi(x,y)
\end{equation}
where $\lambda = \frac{t_x}{t_y}$ and $E$ is normalized by $t_x$. The above equation reflects the tight binding character of the model as it relates $\psi(x,y),  \psi(x \pm a, y) , \psi(x, y \pm a )$. By replacing $x \rightarrow m a $, $ y \rightarrow n b$
and using plane wave solution along the $y$ direction ( as the Hamiltonian is independent of $y$ ) , we have $\psi(x,y) = e^ { i k_y  n } \psi_m$. This leads to a one-dimensional
equation - the  Harper equation\cite{harper}:
\begin{equation}
\psi_{n+1}+\psi_{n-1} + 2  \lambda \cos ( 2 \pi n \phi - k_y) \psi_n = E \psi_n\\
\label{harper}
\end{equation}  

The inherent elegance and beauty of the Harper equation are revealed when we express the corresponding butterfly Hamiltonian using Equation (\ref{tr}). For a square lattice ($\lambda=1$), it is formulated as:
  \begin{equation}
 \hat{H }=  \cos \hat{x} +  \cos \hat{p}, \,\,\,\, \ [\hat{x}, \hat{p}] =  2 \pi i \phi.
 \label{qp}
 \end{equation}
 
 Borrowing Michael Berry's words  - butterfly lives in space of $E$ and (effective) Planck's constant. \\
 
 The derivation above, which begins with a single tight-binding band, corresponds to the limit of strong periodic potential and weak magnetic field. Conversely, in the opposite limit of a weak periodic potential and strong magnetic field, the lattice perturbs the Landau levels. Remarkably, this results in the same Harper equation with $\phi$ replaced by $\frac{1}{\phi}$. This illustrates a strong field-weak field duality, where the two limiting cases—(1) a single Bloch band plus a magnetic field and (2) a single Landau level in a periodic potential—are dual to each other, both described by the Harper's equation. For rational flux $\phi=\frac{p}{q}$, this duality manifests as a "$p \leftrightarrow$ q" symmetry. Throughout this paper, we will witness how this  symmetry magically appears in various equations, including the number-theoretical description of the butterfly.\\
 
Before concluding this sub-section, we briefly mention that the one-dimensional nature of the Harper equation has emerged as a paradigm model for the metal-insulator transition in one dimension when $\phi$ is an irrational number. With $k_y$ as a phase factor, this equation characterizes metallic states for $\lambda < 1$ and localized states for $\lambda > 1$, with $\lambda = 1$ marking the critical point for the onset of the localization transition \cite{KS}, where the spectrum forms a Cantor set. Additionally, this model describes a transition to a strange non-chaotic attractor \cite{SNA}. These aspects related to the Harper equation as a 1D system will not be explored here.\\

 \subsection{ Butterfly Spectrum : Summarizing some key features}
 
We now highlight some of the key features of the butterfly spectrum, which took nearly two decades to observe after Harper initially formulated the butterfly Hamiltonian. Unless otherwise specified, we will assume $\lambda=1$ in all subsequent discussions.

\begin{itemize}
\item The butterfly spectrum is a two-dimensional landscape defined by the magnetic flux interval $0 \le \phi \le 1$ and the energy interval $-4 \le E \le 4$. For each rational value of $\phi = \frac{p}{q}$, there are $q$ bands separated by gaps, where for $q$ even, the central pair of bands touch at $E = 0$.
\item As $\phi$ increases away from any rational value of the flux, such as $\phi_L = \frac{p_L}{q_L}$, the  bands of the spectrum get fragmented and then reforms at  another rational value, $\phi_R = \frac{p_R}{q_R}$ while the gaps separating bands persist.
 It is this reforming of the band with same gaps above and below that is identified as a sub-butterfly.
\item The X-shaped main gaps divide the butterfly graph into central and edge butterflies ( see Fig. (\ref{b4}) ), a division that persists across all levels.  Central and edge butterflies will be referred  as "C-cell" and "E-cell" butterflies. The C-cell butterfly hierarchies with center at flux value $\phi = \frac{p_c}{q_c}$
conserve parity which
we define as even(odd) when $q_c$ is even(odd) while E-cell butterflies do not. They also exhibit distinct topological scaling to be described later. 

 \begin{figure}[htbp] 
 \includegraphics[width = .6 \linewidth,height=.5 \linewidth]{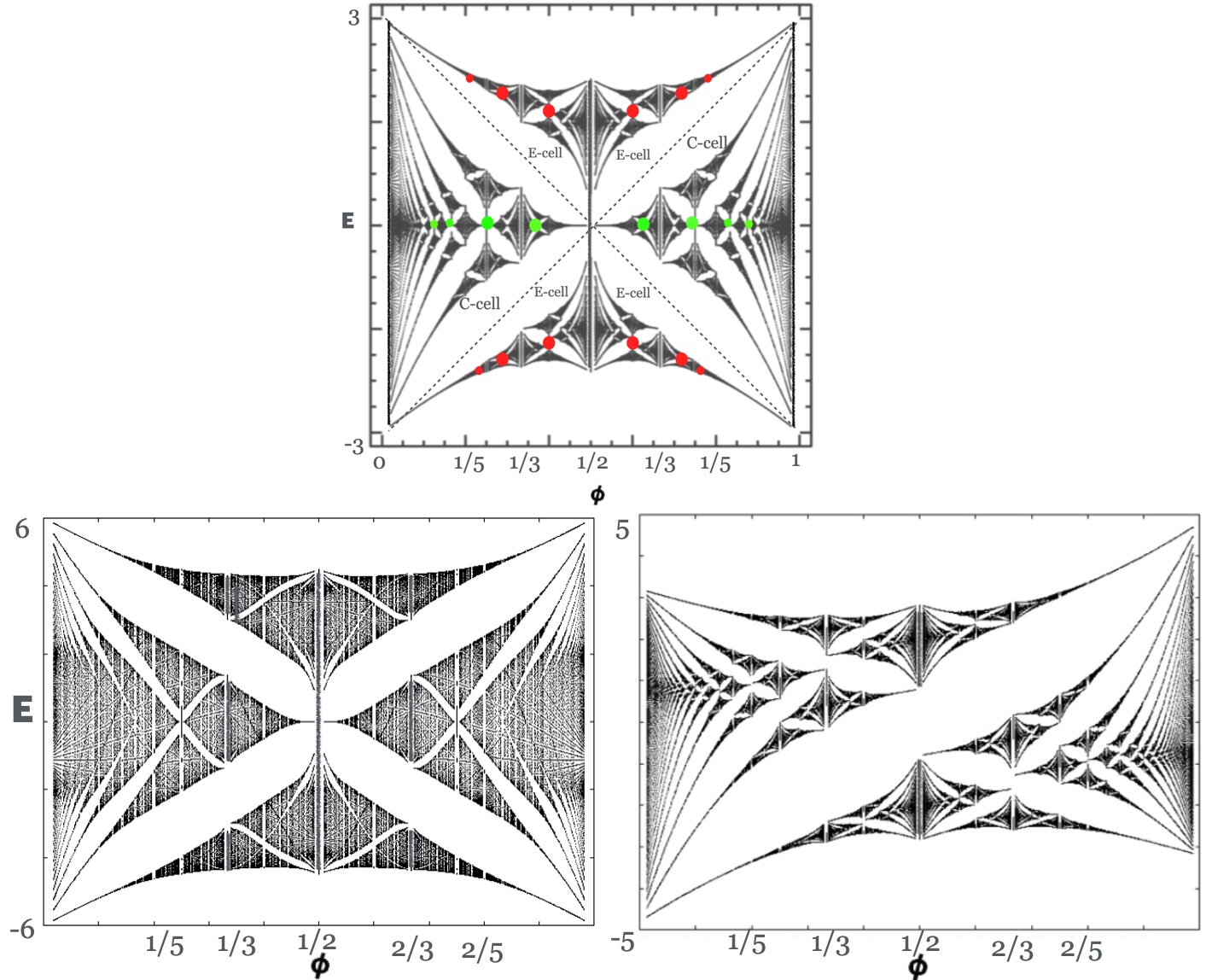} 
\caption{ Upper panel shows the butterfly graph divided in E-cell and C-cell.  Lower left image is for $\lambda=2$ and the right image describes perturbed system
involving small diagonal hopping in the square  lattice.
The diagonal hopping changes the symmetry of the butterfly Hamiltonian, introducing  a gap at flux-$\frac{1}{2}$. These two images illustrate the robustness of ``X-shaped" character of the spectrum. } , 
\label{b4} 
\end{figure}

\item The most striking property of the butterfly spectrum is its self-similarity, as illustrated in Figure (\ref{b1}) ( also see Fig. (\ref{SS})) , where every sub-butterfly is a microcosm of the main butterfly. Understanding aspects of this self-similarity within number theory is a significant focus of this paper and will be elaborated upon later.

\item A key property of every butterfly is that its boundaries and center obey the Farey rule as stated in Equation (\ref{FR}). This forms the basis for relating this fractal to Apollonian gaskets and the Pythagorean tree, discussed in the later part of the paper. The Farey sum rule is Nature's way of incorporating the "strong field - weak field" duality  as described earlier. In other words, the duality relation $\phi \rightarrow \frac{1}{\phi}$ or or $p \leftrightarrow q$, is integrated into the butterfly fractality via the Farey sum. For instance, if $\phi_c$ were simply the arithmetic mean of $\phi_L$ and $\phi_R$, the relationship would not be preserved under this duality.

\item The butterfly graph vividly illustrates the evolution of Bloch bands into Landau levels near $\phi = 0$ and $\phi = 1$. This convergence of Landau levels and Bloch bands, while most pronounced at zero and unit flux, occurs near every rational value of the magnetic flux.  As we will discuss later, this feature of the butterfly graph is embedded in the 2D tessellation of trapezoids and triangles representing the butterfly spectrum, with the tip of the triangle—where the butterfly bandwidth shrinks to zero—symbolizing the Landau level.

\item A notable characteristic of the butterfly graph is that the energy bands are discontinuous functions of the magnetic flux $\phi$, whereas the gaps form continuous channels, except at specific discrete points. This observation, combined with the robustness of these gaps under perturbations (as illustrated in Figure (\ref{b4})), suggests that topological aspects may be at play, given their insensitivity to such perturbations as exemplified in Fig. (\ref{b4}).\\

 Topology, a branch of mathematics, examines the geometric properties of objects that remain unchanged under smooth deformations. By applying the concept of "smooth deformation" to the band structure of a crystalline solid, we can define it as a change in the Hamiltonian that does not close spectral gaps. This concept is explored further in the subsequent section, where the butterfly Hamiltonian is utilized to demonstrate the topological nature of one of the most remarkable phenomena in condensed matter—the quantum Hall effect.
\end{itemize}

\section{ Wannier Diagram-- Butterfly Skeleton emblazoned with integers  }

\begin{figure}[htbp] 
 \includegraphics[width = .75 \linewidth,height=.4 \linewidth]{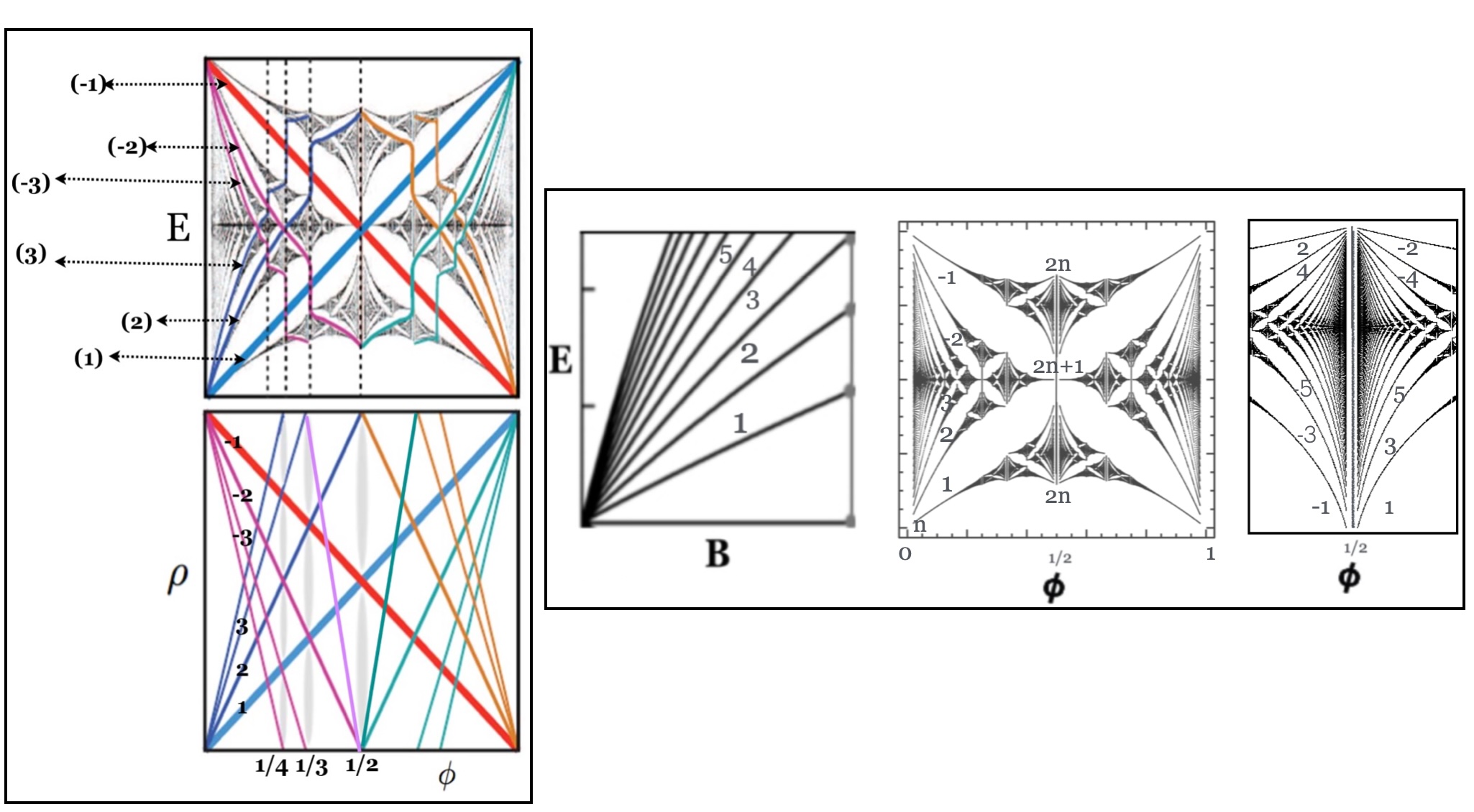} 
 \caption{ The figure illustrates the relationship between the butterfly fractal and the Wannier diagram, with color-coding in the upper graph matching that used in the Wannier diagram below it. This highlights the concept that the Wannier diagram acts as a kind of "skeletal butterfly." In the right panel, images demonstrate how both the Landau levels ( shown in left image ) and the butterfly fractal ( central and right images ) are interwoven with integers. A cluster of  bands resembling Landau levels near flux-$1/2$ are labeled by $2n$ and $2n+1$ where the right image shows a blow near near flux-$1/2$.. These images illustrate that the $n \ne 0$ solutions of the Diophantine equation for a given $\phi$ lie very close to that flux value, emphasizing the important point that near every rational value of $\phi$, Bloch bands degenerate into Landau levels.}
\label{wd}
\end{figure}

The butterfly fractal can be represented more simply, as discovered by Wannier soon after Hofstadter obtained the butterfly spectrum numerically, as shown in Figure (\ref{wd}). Known as the Wannier diagram, the key aspect of this construction was to replace the energies in the butterfly graph with statistical weight, expressed as the integrated density of states of specific segments, following earlier ideas by Ray and Obermair\cite{RO}. For a given value of magnetic flux $\phi$, although the width of the $q$ band in energy generally varies, the statistical weight of each band is $1/q$ (with Wannier considering the total integrated density of states as 1). In other words, instead of using the $(E-\phi)$ variables of the butterfly plot , one constructs the corresponding diagram in 
$(\rho-\phi)$ space, resulting in a geometric fractal - a ``butterfly skeleton" which is isomorphic to the butterfly fractal. 
Wannier's important discovery was that the allowed slanting straight lines are described by the equation:

\begin{equation}
\sigma \phi + \tau = \rho,
\label{D}
\end{equation}

where $(\sigma, \tau)$ are integers. This establishes an isomorphism between the butterfly fractal and the Wannier diagram where the  lines in the Wannier diagram satisfy Eq. (\ref{D}). These straight lines represent the channels formed by the gaps in the butterfly graph, providing a unique labeling of the gaps. Therefore, the Eq. (\ref{D}) is also known as the "gap labeling theorem." However, as Wannier points out, there is a loss of information since the Wannier diagram does not contain energies but only their locations, implying that the Wannier diagram shares the hierarchical pattern of the butterfly fractal. Equation (\ref{D}) was proven by Dana as a general result derived from magnetic translational symmetry\cite{Dana}, meaning it does not depend on specific models and is not limited to strong or weak magnetic fields. For rational flux $\phi = \frac{p}{q}$, there are $(q-1)$ gaps, and for the $r$th gap, $\rho = \frac{r}{q}$. In this case, Eq. (\ref{D}) is a Diophantine equation:
\begin{equation}
\sigma p + \tau q = r .
\label{D1}
\end{equation}

Interestingly, this equation adheres to the "strong field-weak field duality" mentioned earlier, as swapping $p$ and $q$ seamlessly exchanges $\sigma$ and $\tau$.\\

This Diophantine equation  has infinitely many solutions. It is easy to see that if $(\sigma_0, \tau_0)$ is a solution,
then $(\sigma_0+ nq, \tau_0-np)$ with $ n = 0, \pm 1, \pm 2, ...$ are family of solutions. It turns out that for 
 the rectangular lattice, what we want is the smallest possible $\sigma$ ( in absolute value ). Recently, it was shown\cite{SAT21} that this is synonymous with the rule of minimum violation of the butterfly symmetry in the butterfly graph.\\
 
Intriguingly, the  solutions  with $n \ne 0$ find home in the immediate neighborhood of that flux $\frac{p}{q}$ as  the hierarchical set of gaps are clustered 
 in the vicinity of the flux value\cite{book}. To see this,
we explore the neighborhood of $\sigma_0 \phi_0 + \tau_0 = \rho_0$  by substituting $\phi = \phi_0+ \delta \phi$,  $\rho = \rho_0 + \delta \rho$, $\sigma= \sigma_0+\delta \sigma$ and $\tau = \tau_0+\delta \tau$ in Eq. (\ref{D}). Since $\delta \sigma$ and $\delta \tau$ are integers, in  the limit $\delta \phi \rightarrow 0$ and $\delta \rho  \rightarrow 0$, we obtain,
\begin{equation}
 \frac{\delta \sigma}{\delta \tau} = - \frac{q_0}{p_0}, \,\,\,\ {\rm which  \,\ gives}\,\  \delta \sigma = \pm n q_0, \,\ \delta \tau = \mp n p_0.
 \label{delta}
 \end{equation}
 
 Fig. (\ref{wd}) illustrates this result for Landau levels ( showing $\sigma$ values near the cluster of gaps  with $\sigma_0=0$ ) and in the butterfly fractal for flux values $\phi= 0, 1/3, 1/2, 2/3, 1$.
 As we explore in the next section, the integer $\sigma$ represents the quantum number of Hall conductivity and has a topological origin. In essence, the Wannier diagram introduces integers into the butterfly fractal, and these integers are measurable entities in laboratory settings. This establishes a pathway to investigate the fractal aspects of the electron spectrum within a 2D lattice subjected to a magnetic field in experimental environments.\\

\section{Topological aspect of Butterfly Fractal: The Integer  Quantum Hall effect}

In 1980, Klaus von Klitzing's groundbreaking discovery of the quantum Hall effect \cite{QHEK} took the physics community by surprise. This remarkable advancement came nearly a century after Edwin Hall first identified the classical Hall effect in 1879. In the classical Hall effect, when a current flows through a metallic slab in the x-direction and a magnetic field is applied along the z-direction, an induced current appears in the y-direction, known as transverse conductivity,  which we will denote as $G_{xy}$. In contrast, the quantum Hall effect (QHE) revealed that Hall conductivity is quantized  and depends solely on two fundamental constants: $q$ - the charge of the particle, and $h$ - the Planck's constant.
\begin{equation}
G_{xy} =  n \frac{q^2}{h},
\label{SF}
\end{equation}

The quantum Hall effect is remarkable for the extraordinary precision of its quantization—accurate to nearly one part in a billion—regardless of the sample's shape or purity. In 1982, this quantization formula was proven by Streda using a thermodynamic relation known as the Streda formula \cite{Streda}, sometimes referred to as the Widom-Streda formula, which is given by:
\begin{equation}
G_{xy} = q  \frac{\partial N(E)}{\partial B} |_{E= E_F}.
\end{equation}

 Here  $N(E_F)$ is the number of states per unit cell below $E_F$.  While this proof, which also utilized the gap labeling theorem, established the quantization, it did not explain the exceptional precision observed in the phenomenon.\\
 
\begin{figure}[htbp] 
\includegraphics[width = .7\linewidth,height=.4 \linewidth]{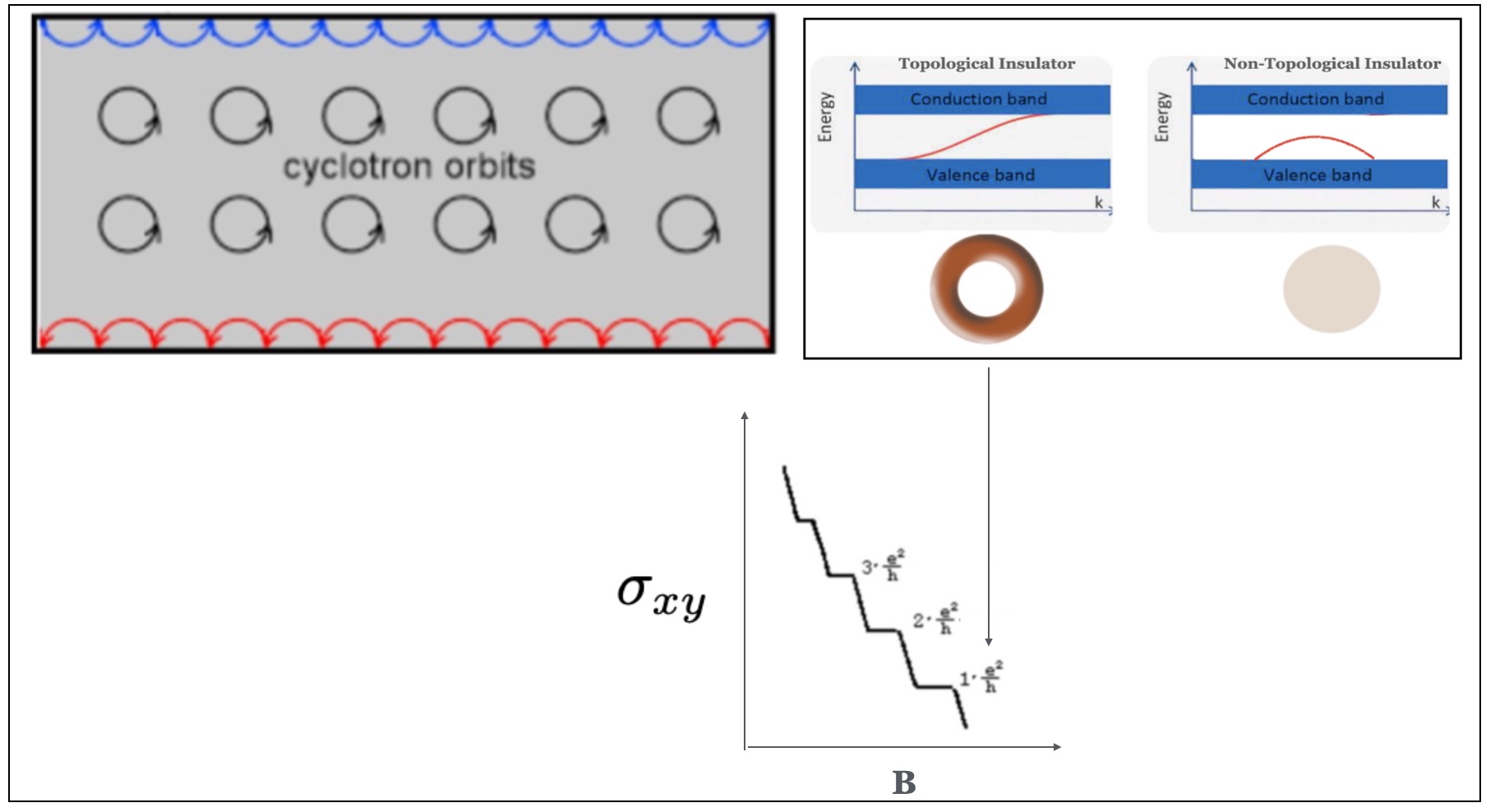} 
\leavevmode \caption{The upper panel illustrates the cyclotron orbits of electrons within a two-dimensional electron gas, along with the bouncing trajectories (depicted in red and blue) of electrons at the sample's upper and lower edges, which are assumed to extend to infinity. These classical trajectories, known as "skipping orbits," consist of a sequence of successive semicircles. Skipping orbits on both edges of the sample carry current, albeit in opposite directions, whereas the circular orbits within the sample do not contribute to the current.
The right panel provides a schematic representation of the band structure for both non-topological and topological insulators. In topological states, the number of edge states corresponds to the quanta of Hall conductivity, which has the topology of a torus (characterized by one hole). In contrast, a trivial insulator is represented by an orange—a shape with no holes.}
\label{B}
\end{figure}

 Later in 1982, Thouless and collaborators utilized the butterfly Hamiltonian model to calculate Hall conductivity using the Kubo formula\cite{TKKN}. They derived Equation (\ref{SF}) and made the groundbreaking discovery that the quantum number $n$ has a topological origin. Remarkably, the quantization of Hall conductivity, calculated in a perfectly non-interacting model, remains unaffected by interactions and impurities due to its topological nature. The full impact of this milestone was realized nearly two decades later when physicists identified other states of matter where topology governs their properties. These novel states, known as topological insulators, are materials that are insulating in the bulk but conductive along the edges. Quantum Hall states are the simplest example of such topological states. Fig. (\ref{B}) highlights the key feature of the quantum Hall states of matter where the bulk of the sample is an insulator and the conduction occurs along the edges.

The quantum Hall effect involves a unique form of quantization, distinct from familiar observables such as energy and angular momentum. Unlike the quantized orbits inside an atom, it does not directly rely on quantum coherence. Instead, two-dimensional electron gases exhibiting this type of quantization are sufficiently cold to maintain quantum coherence, allowing them to be characterized by wave functions that evolve according to the Schr\"{o}dinger equation.\\

{\it The integer quantum numbers of the two-dimensional electron gas are embedded within the wave function.}\\

The emergence of integers in the wave function is attributed to a phenomenon known as quantum anholonomy, where a wave function acquires a phase factor of purely geometrical origin in a cyclic path, referred to as the geometric phase or the Berry phase\cite{BP}. In scenarios where the closed cyclic path forms a closed surface,  this phase factor is an integer multiple of $2\pi$. This occurs precisely when a band is fully filled, corresponding to the spanning of the Brillouin zone (reciprocal space) of a two-dimensional lattice. These integers are topological - known as the Chern numbers. It is important to emphasize that the Berry phase acquired by the wave function results in integers only when a band is completely filled, indicating that the material is an insulator in the bulk\cite{Spinbook}.

  \section{ Farey Tree, Wannier Diagram and 2D Tessellation }
  
  After exploring the quantum Hall effect, where the integers introduced in the butterfly fractal by Wannier come to life as topological entities, we will delve into the number-theoretical aspects of the butterfly diagram. Our journey begins with the Farey tree—a beautiful element of number theory intricately connected to the Wannier diagram \cite{SAT21}.\\

\begin{figure}[htbp] 
 \includegraphics[width = .88 \linewidth,height=.48 \linewidth]{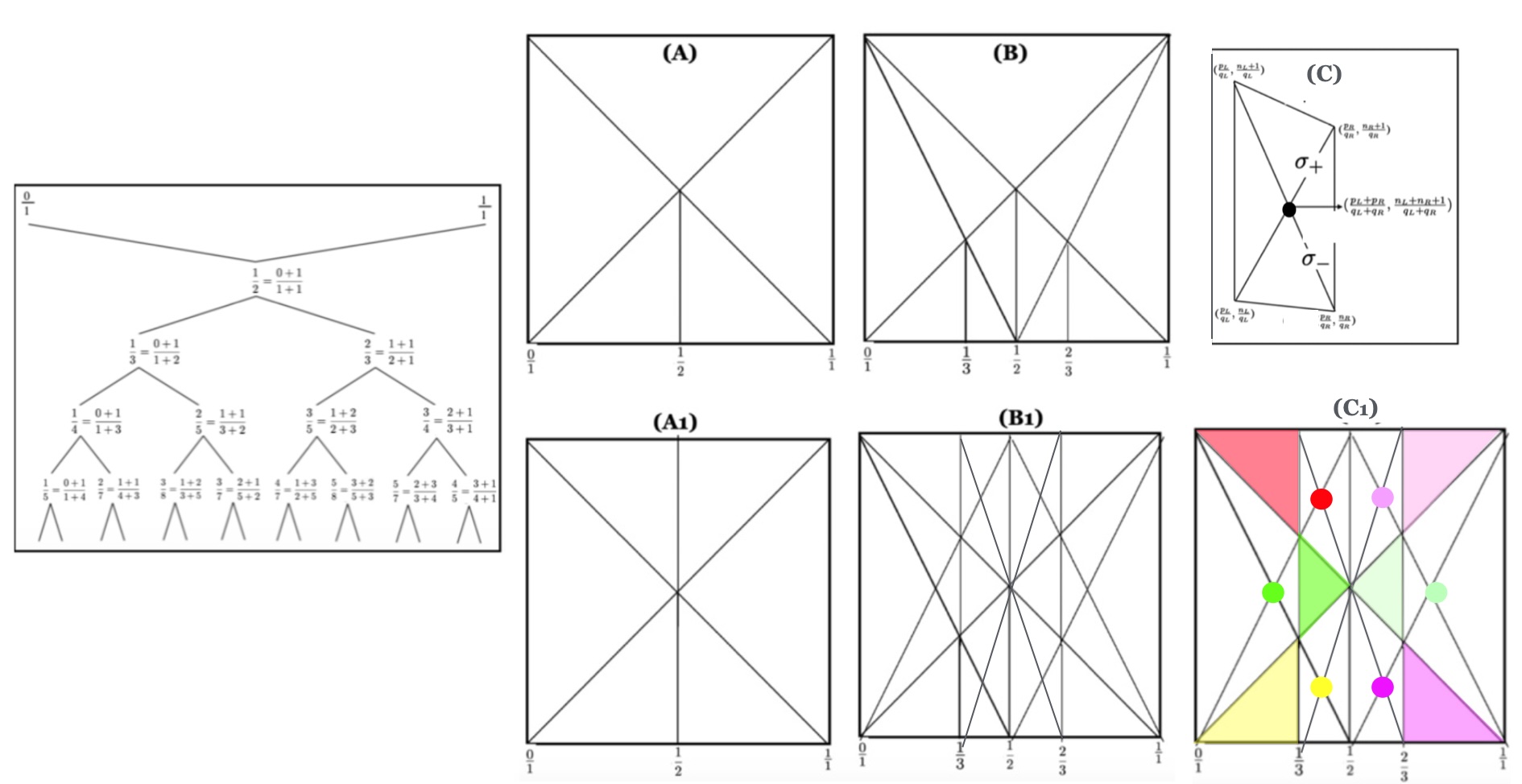} 
 \caption{ The left panel displays the Farey tree constructed using the Farey sum rule (Equation (\ref{FR})). Figures (A-B) illustrate an alternative method for obtaining the Farey tree. The lower panel, $(A_1-B_1)$ depict the corresponding Wannier diagram. Image (C1) emphasizes that (B1) can be interpreted as a 2D tessellation of trapezoids and triangles. Figure (C) demonstrates the Wannier construction of a general trapezoid, where the slopes of the diagonals, denoted as $(\sigma_+, -\sigma_-)$, are integers. The x-coordinate of the trapezoid's center is the sum of the x-coordinates of its two parallel lines, adhering to the Farey sum rule and thus representing the center of the butterfly.}
 \label{FWall}
\end{figure}

Figure (\ref{FWall}) presents two distinct methods for constructing the Farey tree, as highlighted by Hatcher \cite{Hatcher}. Panels (A-B) and their corresponding modifications in (A1-B1) illustrate the relationship between the Farey tree and the Wannier diagram. We will now describe the process in the following steps.

\begin{itemize} 
\item  Start with drawing a unit square  and its diagonals.  Draw a vertical line from the intersection point of the diagonals down to the bottom edge of the square.  Starting with two rational numbers $\frac{0}{1}$ and $\frac{1}{1}$,
 It gives a new rational number $\frac{1}{2}$.  The process generates two trapezoids:
one to the left of $\frac{1}{2}$
with parallel lines at $\frac{0}{1}$ and $\frac{1}{2}$ and another to the right of $\frac{1}{2}$  with parallel lines at $\frac{1}{2}$ and $\frac{1}{1}$.
\item Repeat the above process with each trapezoid: that is draw its diagonals and then draw  vertical lines to the bottom from the intersection point of the diagonals of the trapezoids. This gives the Farey fraction $\frac{1}{3}$ and $\frac{2}{3}$.  Vertical line from each of the fraction generates two new trapezoids, one to the left  and the other to the right of that fraction. 
\item  Continue this process: with each new trapezoid, draw its diagonals and the vertical line from the point of intersection of the diagonals.
 This will generate all rationals because the vertical lines from the diagonals of the trapezoid  formed by two parallel lines at $\frac{p_L}{q_L}$ and $\frac{p_R}{q_R}$ meet the bottom edge at $\frac{p_L+p_R}{q_L+q_R}$ - the Farey sum of the two fractions. This is  shown in panel (D).
\item  In general, for every trapezoid so formed with two parallel lines at  friendly fractions $\frac{p_L}{q_L}$ and $\frac{p_R}{q_R}$, the y-coordinates of the upper left and the upper right corners of the trapezoid are $\frac{1}{q_L}$ and $\frac{1}{q_L}$. The coordinate of the intersection of the diagonals is $( \frac{p_L+p_R}{q_L+q_R}, \frac{1}{q_L+q_R})$. By induction, this proves that the length of every vertical  line of the trapezoid  at
fraction $\frac{p}{q}$ is $\frac{1}{q}$. 
\end{itemize}

In summary, the Wannier diagram emerges from the geometric construction of the Farey tree (Figure (\ref{FWall})) through two steps. First, all vertical lines are extended to the upper edge of the unit square, which corresponds to the $\rho=1$ line of the Wannier diagram. Second, new lines are introduced to ensure that the entire configuration is symmetrical about $\rho=1/2$. This implies that the density $\rho$ as a function of $\phi$ satisfies the condition $\rho(\phi)= \rho(1-\phi)$.
A key point to note is that all the slanting lines in Figure (\ref{FWall}) have integer slopes and integer intercepts when the parallel lines of the trapezoid are at friendly fractions, with the height of each parallel line at  $\frac{p}{q}$ being $\frac{1}{q}$. For example, in a general trapezoid shown in panel (D), the slopes of the two diagonals, $(\sigma_+, -\sigma_-)$,  are given by:

\begin{equation}
\sigma_+  = | n_L q_R - (n_R+1) q_L | ,\,\,\,\ \sigma_- = | (n_L+1)  q_R - n_R q_L]|.\\
\end{equation}
\bigskip\par

Now we address the key question: {\it what lines in the Wannier diagram represent a butterfly ?}\\

In order for a trapezoid to represent a butterfly,
that is to form a butterfly skeleton, $(\sigma_+, \sigma_-)$ must correspond to the smallest possible solution of Eq. (\ref{D1}).  This corresponds to minimum value of
  $\Delta \sigma$\cite{SAT21}, which is defined as:
\begin{equation}
\Delta \sigma = \sigma_+ - \sigma_- .
\label{asy}
\end{equation}

Nature's choice of using smallest possible solutions of Diophantine equation ((Eq. (\ref{D1})) is equivalent to smallest possible values of $\Delta \sigma$.
Since a butterfly with $\Delta \sigma=0$ have mirror symmetry about the x-axis, $\Delta \sigma$ is a measure of asymmetry of a butterfly.
Therefore,  within a rectangular strip at flux values $(\frac{p_L}{q_L}, \frac{p_R}{q_R})$, trapezoids with minimum violation of mirror symmetry form butterflies in the butterfly fractal\cite{SAT21}. \\

\subsection{ Labeling a butterfly with three integers}

{\it How many integers are needed to label a butterfly uniquely ? }\\

We will now argue that just three integers are sufficient to label a butterfly uniquely. 

\begin{itemize}
\item The triplet $\Big[\frac{p_L}{q_L}, \frac{p_c}{q_c}, \frac{p_R}{q_R}\Big]$ 
 respectively represent the magnetic flux values at the left, at the center  and at the right boundary of a butterfly in the butterfly graph.  In addition to Farey sum rule, (Eq. (\ref{FR})),
 these integers satisfy:
 \begin{equation}
p_x q_y - p_y q_x = \pm 1,  x \ne y
\label{F}
\end{equation}
 Given a pair $(q_R, q_L)$, the corresponding  pair $(p_R, p_L)$ is uniquely determined from the  above Diophantine equation Eq. (\ref{F}) as both $(p_L, q_L)$ and $(p_R,q_R)$ are relatively prime and the fractions $\frac{p_L}{q_L}$ and $\frac{p_R}{q_R}$ are less than unity.
 \item Butterfly is characterized by topological quantum numbers, called the Chern numbers\cite{TKKN}. These are the band Chern number  $N$ and the Chern numbers of two gaps ( that form X-shape),
 labelled $(\sigma_+, - \sigma_-)$ where  $\sigma_{\pm } > 0 $.   In the geometrical representation of the butterfly as  a trapezoid, $\sigma_+, -\sigma_- $ are the slopes of the two diagonals of the trapezoid.
 The gap Cherns $\sigma$ and the band Chern $N$  satisfy linear Diophantine equations\cite{TKKN, Dana}:
 \begin{equation}
 p \sigma + q \tau = r ,\,\,\,\ pN + qM = 1,
 \end{equation}
 Here $r$ labels $r^{th}$ gap of the energy spectrum at magnetic flux $\frac{p}{q}$ and  $\tau$ and $M$ are other integers.
\item Integers described above satisfy the following equations:
\begin{eqnarray}
q_L - q_R  =  N, \quad 
q_L + q_R  =  \sigma_+ + \sigma_- 
\end{eqnarray}
\item  In view of  above, a close look at the general trapezoid represented in Fig. (\ref{FWall}), it is clear that the trapezoid - the butterfly skeleton is uniquely determined by three integers.
We can choose them to be $( q_R, q_L, \sigma_+)$ or $( N, \sigma_+, \sigma_-)$ or $( q_R, q_L,  \Delta \sigma)$  . This is an important aspect of
the butterfly  fractal as its skeleton representation  {\it solely determined by topology }.
\end{itemize}
   \section{ ``Butterfly with a tail" as the Building block of the the Butterfly fractal}
 
 \begin{figure}[htbp] 
\includegraphics[width = .65 \linewidth,height=.27 \linewidth]{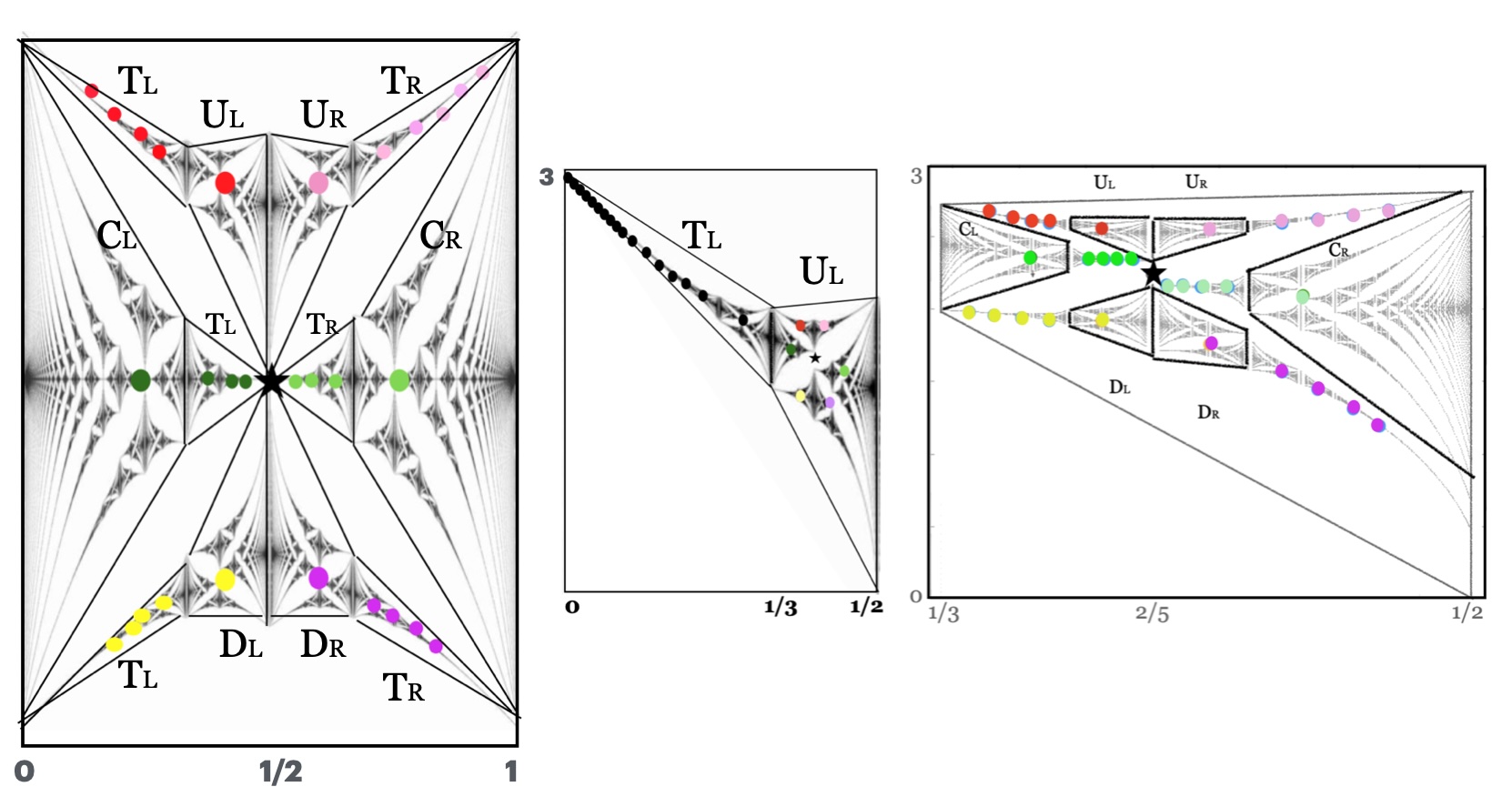} 
\caption{ Figure shows butterfly attached to a tail and its six baby butterflies.  This representation is valid for the main butterfly ( left), its baby butterfly $U_L$ 
(middle) and the blow up of the baby butterfly ( right panel).}
 \label{tail}
\end{figure}

Figure (\ref{b2}) provides a foundational guide for constructing the butterfly graph. The concept of the butterfly with a tail as a fundamental building block is further detailed in Figure (\ref{tail}). The recursive pattern features a parent butterfly generating six baby butterflies, each accompanied by a tail—a chain of butterflies with progressively decreasing sizes, culminating at a limit point. In the two-dimensional tessellation depicted in panel $C_1$ of Figure (\ref{FWall}), trapezoids represent the butterflies, while a triangle signifies an infinite chain of butterflies with shrinking magnetic flux intervals and energy bands. The tips of the triangles, where the butterfly converges to a point, symbolize the union of the Bloch bands with the Landau levels.\\

The mathematical framework involves a total of eight generators: six for generating the six baby butterflies, denoted as  $(C_L, C_R, U_L, U_R, D_L, D_R)$, and two for generating the tail, denoted as $(T_L, T_R)$. Here, $C_L$ and $C_R$ are the generators for the central butterfly, while $(U_L, U_R, D_L, D_R)$ are the generators for the upper and lower edge butterflies that appear to the left and right of the parent center. The operators $T_L$ and $T_R$ generate the left and right tails, respectively. Since $(U_L, D_L)$ and $(U_R, D_R)$, which generate the E-cell butterflies, share the flux intervals, they are collectively represented as $E_L$ and $E_R$, respectively.\\

The following two sub-sections describe the magnetic flux $\phi$ recursions and the recursions for the gap Chern numbers $(\sigma_{\pm})$ underlying the butterfly tree. Interestingly, the $\phi$ recursions are decoupled from $\sigma$-recursions. However, recursions for $\sigma$ are coupled with $( q_L, q_R) $. As described later,  the $\phi$ recursions can also be derived from number theory.

\subsection{ Recursion Relation for the Magnetic Flux}
The $\phi$ recursions, derived through careful examination of the butterfly graph, are presented below. They illustrate how Farey sequences form the fundamental components of the butterfly tree, with Farey neighbors playing a crucial role. There are two families of Farey neighbors : 
  $ (\phi_L, \phi_R, \phi_+)$ and  $ (\phi_L, \phi_R, \phi_-)$ where $\phi_{\pm} = |\frac{p_L \pm p_R }{q_l \pm q_R}|$. Known as friendly triplets,  any two members of  each of these triplets
  satisfy Eq. (\ref{F}).
\bigskip\par\noindent
\begin{eqnarray*}
\label{l1}
E_L \equiv (U_L, D_L)   :     \frac{p_R}{q_R}(l+1)   & =  &  \frac{p_{+}}{q_{+}}(l) ,\,\,\,\  \frac{p_L}{q_L}(l+1) = \frac{p_L}{q_L} \bigoplus \frac{p_+}{q_+} (l)  \\
\label{l2}
E_R \equiv (U_R, D_R)   :    \frac{p_L}{q_L}(l+1) &  =  & \frac{p_{+}}{q_{+}}(l)  ,\,\,\,\   \frac{p_R}{q_R}(l+1) = \frac{p_R}{q_R}  \bigoplus \frac{p_+}{q_+}(l) \\
\\
C_L   :   \frac{p_L}{q_L}(l+1)  & = & \frac{p_L}{q_L}(l)  ,\,\,\,\   \frac{p_R}{q_R}(l+1) = \frac{p_L}{q_L} \bigoplus \frac{p_+}{q_+}(l) \\
C_R   :   \frac{p_R}{q_R}(l+1) &  =  & \frac{p_R}{q_R}(l)  ,\,\,\,\   \frac{p_L}{q_L}(l+1) = \frac{p_R}{q_R} \bigoplus \frac{p_+}{q_+}(l)\\ 
\\
T_{L}   :     \frac{p_R}{q_R}(l+1)  & =  & \frac{p_L}{q_L}(l)  ,\,\,\,\  \frac{p_L}{q_L}(l+1) =  \frac{p_L}{q_L} \bigoplus \frac{p_-}{q_-}(l) \\
T_{R}   :    \frac{p_L}{q_L}(l+1) &  =  & \frac{p_R}{q_R}(l)  ,\,\,\,\   \frac{p_R}{q_R}(l+1) = \frac{p_R}{q_R}  \bigoplus \frac{p_-} {q_-}(l) 
\end{eqnarray*}

These recursions can also be written as $2 \times 2$ matrices that act  $\left( \begin{array}{c}  q_L  \\  q_R   \\  \end{array}\right)$ and $\left( \begin{array}{c}  p_L  \\  p_R   \\  \end{array}\right)$.

\begin{equation}
E_L =  \left( \begin{array}{cc} 1   & 1  \\  1 & 2   \\  \end{array}\right), \quad  E_R =  \left( \begin{array}{cc} 2   & 1  \\  1 & 1   \\  \end{array}\right) ,\,\ c_L =  \left( \begin{array}{cc} 1   & 2  \\  0 & 1   \\  \end{array}\right),\,\ c_R =  \left( \begin{array}{cc} 1   & 0  \\  2 & 1   \\  \end{array}\right), \,\  \tau_{R}=  \left( \begin{array}{cc} 2   & -1  \\  1 & 0   \\  \end{array}\right), \,\  \tau_{L} =\tau^{-1}_{ R},
\label{M2} \end{equation}

These four generators - elements of the group $SL(2,Z))$, distinctly differentiate between the E-cell butterflies and the C-cell recursions. Known as Arnold’s cat map, the generators for the E-cell butterflies have a trace of $3$, while those for the C-cell butterflies have a trace of $2$. A more nuanced characterization of this difference becomes evident when examining the fixed points associated with these maps. To identify these fixed points, we decouple the equations, resulting in two-term recursions for $q_x$ and explore the asymptotic limit for the ratio of these variables. Below, we demonstrate this dichotomy for $U_L$ and $C_L$.

\begin{eqnarray}
\label{decouple}
E_L  & :  & \,\ q_x(l+2)-3q_x(l+1)+q_x = 0, \quad \,\,\ 
   \lim_{l  \rightarrow   \infty} \frac{q_x(l+1)}{q_x(l)} = \frac{3 \pm \sqrt{5}}{2} \nonumber \\
   \\
 C_L  &   :  &  \,\ q_L(l+1)-q_L(l)=0,  \quad \,\,\,\,\,\,\,\,\,\,\,\,\,\,\,\,\,\,\,\,\    \lim_{l  \rightarrow   \infty} \frac{q_x(l+1)}{q_x(l)} = 1
\end{eqnarray}

As described below, C-cell butterflies exhibit non-trivial scaling as we explore  the relationship between any two sub-butterfly as described in section VII.
The dichotomy in the E-cell and C-cell butterflies is actually characterized by the scaling of the topological integers characterized by $\Delta \sigma$ (Eq. (\ref{asy})) whose recursion relation is described 
below.

\subsection{ Recursion Relations for the Gap Chern Numbers } 

The two  butterflies  labeled $(U_L, D_L)$ or  $(U_R, D_R)$  share the same flux intervals and hence also the band Chern numbers.  However, they are topologically distinct 
, characterized by different values of  $(\sigma_+, \sigma_-)$ - that is the Chern numbers of the two main gaps of the butterflies. To obtain a unique label for each butterfly, we need both the flux interval as well as the gap Chern numbers\cite{SAT25}. \\

To obtain the recursion relations for $(\sigma_+, \sigma_-)$, we make use of Eq. (\ref{delta}) that the Chern numbers of the hierarchical set of gaps near a rational flux value say $\frac{p_0}{q_0}$ are $\sigma_0+ nq_0$ where $\sigma_0$ is the Chern number of the parent gap. 
Using the equations (\ref{l1}) and (\ref{l2}), along with  the identity, $\sigma_+ + \sigma_- = q_c$, we can derive the renormalization trajectories of the
Chern numbers $( \sigma_+, \sigma_-)$ and hence $\Delta \sigma$. 
\begin{eqnarray}
\label{R8}
U_L : \,\  \sigma_+ (l+1) & = &  \sigma_+(l) + q_L(l),  \quad   \sigma_- (l+1)  =   \sigma_-(l) + q_+(l), \quad  \Delta \sigma(l+1) = \Delta \sigma(l)-q_R(l) \\
U_R : \,\   \sigma_+ (l+1) & = &   \sigma_+(l) + q_+(l) ,   \quad   \sigma_- (l+1)  =   \sigma_-(l) + q_R(l), \quad  \Delta \sigma(l+1) = \Delta \sigma(l)+q_L(l) \\
D_L : \,\  \sigma_+ (l+1)  &  =   &   \sigma_+(l)+ q_+(l) ,  \quad   \sigma_- (l+1)  =   \sigma_-(l) + q_L(l), \quad  \Delta \sigma(l+1) = \Delta \sigma(l)+q_R(l) \\
D_R : \,\  \sigma_+  (l+1)   & =  &   \sigma_+(l) + q_R (l),   \quad   \sigma_- (l+1)  =   \sigma_-(l) + q_+(l), \quad  \Delta \sigma(l+1) = \Delta \sigma(l)-q_L(l) \\ 
\nonumber \\
C_L : \,\   \sigma_+ (l+1) & = &   \sigma_+(l) + q_L(l) ,   \quad   \sigma_- (l+1)  =   \sigma_-(l) + q_L(l), \quad  \Delta \sigma(l+1) = \Delta \sigma(l) \\
C_R : \,\  \sigma_+  (l+1) & =  &  \sigma_+(l) + q_R (l), \quad   \sigma_- (l+1)  =   \sigma_-(l) + q_R(l), \quad  \Delta \sigma(l+1) = \Delta \sigma(l) \\
\nonumber \\
T_L : \,\  \sigma_+  (l+1) & =  &  \sigma_+(l) + q_- (l), \quad   \sigma_- (l+1)  =   \sigma_-(l) + q_-(l), \quad  \Delta \sigma(l+1) = \Delta \sigma(l) \\
T_R : \,\  \sigma_+  (l+1) & =  &  \sigma_+(l) + q_- (l), \quad   \sigma_- (l+1)  =   \sigma_-(l) + q_-(l), \quad  \Delta \sigma(l+1) = \Delta \sigma(l) 
\label{RR8}
\end{eqnarray}

Combined with the recursions for $q_R,q_L$, recursions for three integers  $(q_R, q_L, \Delta \sigma)$ that provide unique labeling of a butterfly are given by: 

\begin{eqnarray}
\label{E1}
U_L  & =  & \left( \begin{array}{ccc}   1 & 1 & 0  \\   1 & 2 & 0 \\    -1 & 0 & 1  \\ 
\end{array}\right), D_L   =   \left( \begin{array}{ccc}  1 & 1 & 0  \\   1 & 2 & 0  \\  1 & 0 & 1 \\   
\end{array}\right),  U_R   =   \left( \begin{array}{ccc}    2 & 1 & 0 \\  1 & 1 & 0  \\    0 & 1 & 1 \\  
\end{array}\right), D_R = \left( \begin{array}{ccc}   2 & 1 & 0  \\  1 & 1 & 0  \\   0 & -1 & 1 \\    
\end{array}\right)\\
C_L  & =  & \left( \begin{array}{ccc}   1 & 2 & 0  \\   0 & 1 & 0  \\    0 & 0 & 1\\ 
\end{array}\right), C_R = \left( \begin{array}{ccc}  1 & 0 & 0  \\   2 & 1 & 0 \\    0 & 0 & 1  \\  \end{array}\right),  T_{L}  =   \left( \begin{array}{ccc} 0 & 1 & 0  \\   -1 & 2 & 0  \\    0 & 0 & 1 \\  \end{array}\right) , 
T_{R} = \left( \begin{array}{ccc}   2 & -1 & 0  \\  1 & 0 & 0 \\    0 & 0 &  1 \\
\end{array}\right)
\label{E2}
\end{eqnarray}
 \begin{figure}[htbp] 
\includegraphics[width = .5 \linewidth,height=.38 \linewidth]{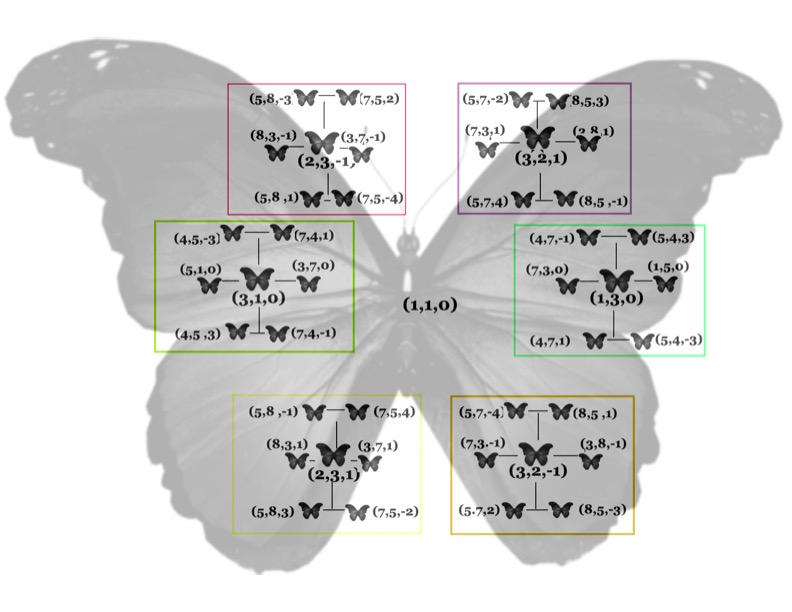} 
\leavevmode \caption{ Three generations of butterfly tree.  Starting with the parent butterfly  labelled as $(1,1,0)$,  figure shows labeling of its
six babies . Each of these six
babies produce their own sextuplets, that is $18$ grand babies of  $(1,1,0)$. Not
displayed in the figure are the butterfly tails.   }
\label{tree}
\end{figure}

The tails of the butterflies are obtained by  repeated applications of $T_L$ ( for tail attached to the left of the butterfly) and $T_R$ ( for tail attached to the right of the butterfly).
  Figure (\ref{tree}) displays the tree  of the triplet $(q_R, q_L, \Delta \sigma)$- the building blocks of butterfly fractal.
For simplicity, parts of the tree corresponding to the tails are not shown.
To summarize the iterative process, one starts with a butterfly, which we refer as parent butterfly and use $( C_L, C_R, U_L, U_R, D_L, D_R)$ generators to produce the sextuplets. One then uses $T_{L}$ for $(C_L, U_L, D_L)$ generated butterflies and $T_{R}$ for $(C_R, U_R, D_R )$  to generate their tails.
The process is repeated ad infinitum for each of the six butterflies as well as for the butterflies in the chains  consisting of infinity of of butterflies that form the tails.\\

\subsection{ Relation to ${\bf SL(2,Z) \times Z} $ group}

 \begin{figure}[htbp] 
\includegraphics[width = .3\linewidth,height=.28 \linewidth]{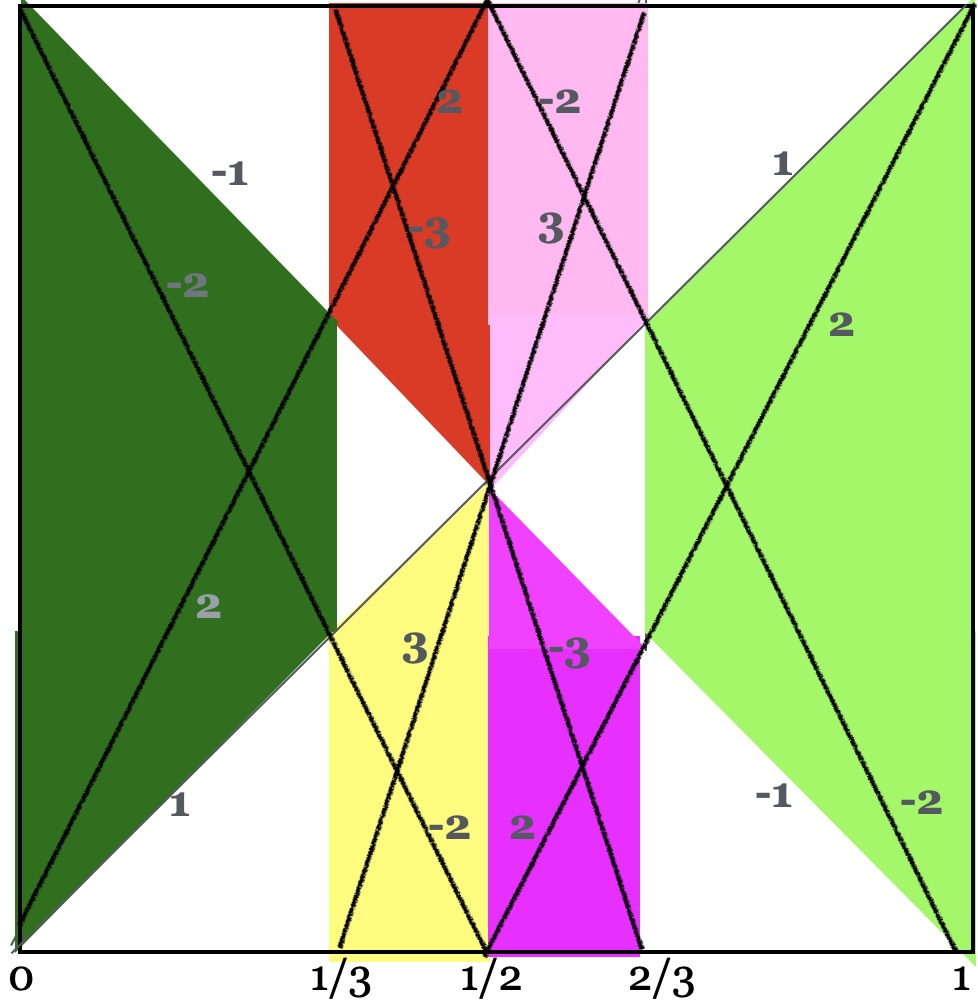} 
\leavevmode \caption{  The first generation of the skeleton graph , illustrating
how the six butterfly babies (shown with six distinct colors) share the
topological quantum numbers $(\sigma_+, \sigma_-)$. Shown in black, these integers are the
slopes of the diagonals of the trapezoid which due to the Farey relation are always
integers. }
\label{couple}
\end{figure}

As described above, C-cell butterflies and  the butterflies in the tail are described by matrices that are block diagonal and hence belong  to the group $SL(2,Z) \times Z$. However, the E-cell butterflies lack this simple characterization. It turns out that by a different parametrization for the E-cell  butterflies facilitates  their generators to be also part of the $SL(2,Z)\times Z$ group. To achieve that, we are guided by 
Fig (\ref{couple}) , showing {\it sharing} of $\sigma_{\pm}$ and $(q_L,q_R)$ among the six sibling butterflies. The shared $\sigma_{\pm}$ are stated explicitly below.
\begin{eqnarray}
\label{T1}
\sigma_+(U_L) & = &  \sigma_+(C_L) ,\,\,\,\ \sigma_-(D_L)  =   \sigma_-(C_L)\\
\sigma_+(D_R) & = &  \sigma_+(C_R) ,\,\,\,\ \sigma_-(U_R)  =   \sigma_-(C_R) \\
\label{T2}
\sigma_{-}(U_L) & = &  \sigma_{-}(D_R) ,\,\,\,\ \sigma_{+}(U_R)  =   \sigma_{+}(D_L)
\end{eqnarray}
We will now leverage the sharing characteristics of the E-cell butterfly siblings to construct a simple and elegant formulation of the recursions for E-cell butterflies. 
The key idea is to  replace $(q_R, q_L, \sigma_+)$ with $(q_s, q_{ns}, \sigma_s)$, where $q_s$ and $q_{ns}$ respectively represent the "shared" and "not shared" values of $q$ between the left and right edge butterflies, and $\sigma_s$ is the shared value of $\sigma$ between the upper left (right) and lower right (left) edge butterflies. For example, $q_s = q_R$ for $E_L$ and $q_s = q_L$ for $E_R$; $q_{ns} = q_L$ for $E_L$ and $q_{ns} = q_R$ for $E_R$. Similarly, $\sigma_s = \sigma_-$ for $(U_L, D_R)$ and $\sigma_s = \sigma_+$ for $(U_R, D_L)$. It is easy to see that all four generators $U_L, U_R, D_L, D_R$, as displayed in Equation (\ref{T2}), become the elements of to $SL(2,Z) \times Z$, where the $SL(2,Z)$ matrix is represented by a single matrix $E$ and $Z$ corresponds to the values of $\sigma_s - q_{ns}$, which we denote as $\alpha$. For the C-cell butterflies, we define $(q_s, q_{ns})$ as the $q$ values that the butterfly shares with the edge sibling, and $\alpha = \Delta \sigma$. Thus, the E-cell and C-cell butterflies are defined by just two generators, denoted as $\mathcal E$ and $\mathcal C$.

 \begin{eqnarray*}
 {\rm E-cell}  & : &  \mathcal{E} =  \left( \begin{array}{cc} 1   & 1  \\  1 & 2   \\  \end{array}\right) \,\,\ ,\,\,\  \alpha = \sigma_s - q_{ns} \nonumber \\
 \\
  {\rm C-cell}  & : &  \mathcal{C} =  \left( \begin{array}{cc} 1   & 2  \\  0 & 1   \\  \end{array}\right) \,\,\ ,\,\,\  \alpha = \Delta \sigma  \\
  \end{eqnarray*}
  where,
  \begin{equation}
  \mathcal{E}  \left( \begin{array}{c} q_s(l)  \\  q_{ns}(l)   \\  \end{array}\right) =  \left( \begin{array}{c} q_s(l+1)  \\  q_{ns}(l+1)   \\  \end{array}\right), \quad
   \mathcal{C}  \left( \begin{array}{c} q_s(l)  \\  q_{ns}(l)   \\  \end{array}\right) =  \left( \begin{array}{c} q_s(l+1)  \\  q_{ns}(l+1)   \\  \end{array}\right)
  \end{equation}
\section{ Self-similarity of the Butterfly Fractal  }
\begin{figure}[htbp] 
\includegraphics[width = .5\linewidth,height=0.35\linewidth]{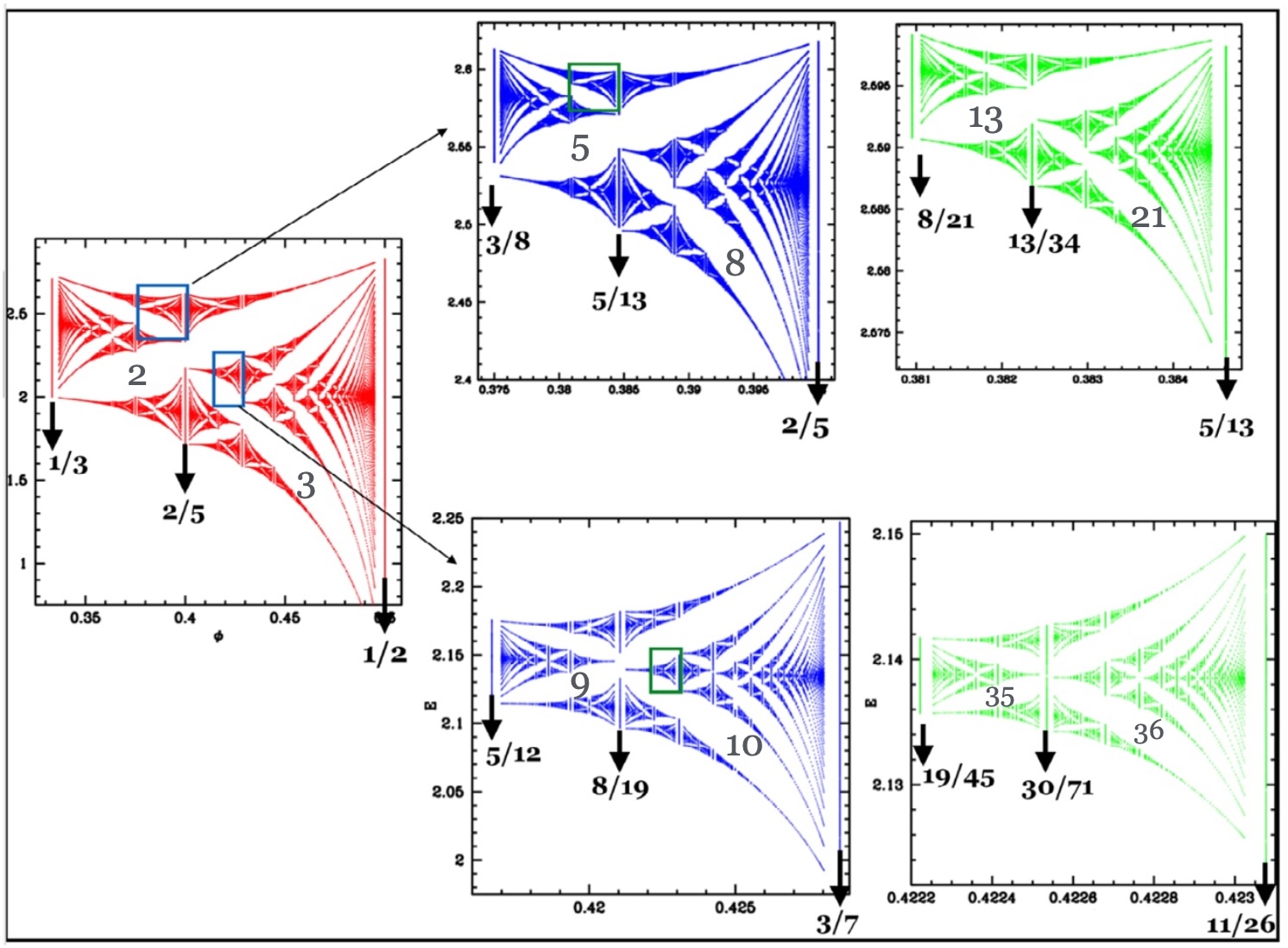}
\leavevmode \caption{ Illustrating three generations of two distinct nested sequences of sub-images where the upper set of images show E-cell nesting and the lower sets are for a C-cell.
Starting with the main butterfly, the E-cell hierarchy  ( upper images ) is obtained by repeated application of $U_L$. The C-cell hierarchy ( lower images ) are obtained
first using $U_L$  and then repeated application of $T_LC_R$.  With flux values and topological numbers labeled, we can verify asymptotic scaling $\zeta$ in these two cases 
which are $\frac{3+\sqrt{5}}{2}$ and $2+\sqrt{3}$  respectively. These scale factors are   the eigenvalue of $U_L$ for upper images and eigenvalue of  $T_LC_R$  for the lower images.
Alternatively, these scale factors can be determined by  Eq. (\ref{zeta1}). }
\label{SS}
\end{figure}

"Worlds within worlds" aptly describes the self-similar characteristics of the butterfly fractal, as each sub-butterfly serves as a microcosm of the main butterfly. This concept is illustrated in 
Figures (\ref{SS}) ( also see Fig. (\ref{b1}) ).  

Each butterfly hierarchy is characterized by a scaling exponent $\zeta$. It turns out that a single exponent determines the scaling for all  the integers associated with the butterfly:
\begin{eqnarray}
\label{scal}
{\rm E  \,\ \& \,\ C \,\ Cells}  :  q_x(l)  &  \sim  &  \zeta^l ,\,\,\,\ p_x(l)  \sim  \zeta^l ,\,\,\,\  \sigma_{\pm}  \sim  \zeta^l  \\
\nonumber \\
\label{scal1}
{\rm E-Cell}:  & \,\  & \Delta \sigma(l) \sim \zeta^l , \quad  \,\ {\rm C-Cell}: \,\ \Delta \sigma(l) \,\  =  {\rm constant}
\end{eqnarray}

The scaling factor $\zeta$ is given by the product of generators whose repeated applications determine the magnetic flux values and the topological integers associated with nested set of butterflies. 
The RG scheme described in earlier study\cite{SW} determines a complete set of $\zeta$ values that describe the entire butterfly fractal.  Here, starting with the parent butterfly as the main butterfly, choose a sub-butterfly characterized by $( p^*_L, q^*_L, M^*, N^* )$. The RG equation to describe self-similar recursion relation is:
\begin{equation}
\phi_x (l+1) = \frac{M^* \phi_x(l) + p^*_L}{-N^* \phi_x(l) + q^*_L} 
\label{R1}
\end{equation}
where $ x = L,c, R $. In other words, given $[\phi_L(1), \phi_c(1), \phi_R(1)]=
[\frac{0}{1}, \frac{1}{2}, \frac{1}{1} ] $,  the boundaries and the centers of the sub-images are the iterates of Eq. (\ref{R1}).
The RG equations that describe recursions of  butterfly integers $(p_L, q_L, p_R, q_R, N, M )$ are discussed in detail in previous studies\cite{SW, SAT20} where they can be cast in various forms. One of the elegant representations  involves writing recursions as a $2 \times 2$ matrix. Define  matrices  $C^*_L,  C^*_R$  and $A_x(l)$ ( x = L, R) as:
\begin{equation}
\label{eq: 4.4a}
{\bf A}_{\rm x} (l)=
\left(\begin{array}{cc}
q_{\rm x}(l) & p_{\rm x} (l)\cr
-N(l) & M(l)
\end{array}\right),\,\
{\bf C}^*_{\rm L}=
\left(\begin{array}{cc}
 q^*_{\rm L} &  p^*_{\rm L} \cr
- N^* &  M^*
\end{array}\right)
,\,\ 
{\bf C}^*_{\rm R}=
\left(\begin{array}{cc}
 q^*_{\rm R} &  p^*_{\rm R}- q^*_{\rm R}\cr
- N^* &  M^*+ N^*
\end{array}\right)
\ .
\end{equation}

The RG equations for the left and the right flux boundaries are  given by:
\begin{equation}
\label{eq: 4.4b}
{\bf A}_{\rm L} (l+1)={\bf C}^*_{\rm L} {\bf A}_{\rm L}(l) ,\,\,\ {\bf A}_{\rm R} (l+1)={\bf C}^*_{\rm R} {\bf A}_{\rm L}(l) 
\ .
\end{equation}

The  scaling exponents $\zeta$ associated with Eq. (\ref{R1}) are the eigenvalue of the matrix $C_L^*$ or $C^*_R$. Using the identity $p^* N^*+q^* M^* =1$, the scaling exponents are given by:
\begin{equation}
\zeta = \frac{( n^*+2)}{2}\pm\sqrt{\left(\frac{n^*+2}{2}\right)^2-1} , \,\,\ \quad n^* = q^*_L + p^*_R-p^*_L-2
 \label{zeta1}
 \end{equation}
Expressed as a  continued fraction expansion, these quadratic irrationals  are given by $\zeta$:
 \begin{eqnarray}
\zeta = [n^*+1; \overline{ 1, n^*} ]& \equiv &  
( n^*+1)+ \cfrac{1}{1
          + \cfrac{1}{n^*
          +\cfrac{1}{1
          +\cfrac{1}{n^*
          + \cfrac{1}{1 + \cfrac{1}{n^*.....} } } }}}
          \label{cont}
          \end{eqnarray}
          
Thus, each self-similar butterfly hierarchy in the butterfly graph is labeled by an integer $n^*$. It is quite remarkable that nature employs this special class of quadratic irrationals to describe the self-similar characteristics of the butterfly graph. It is important to note that $n^*$ does not provide a unique characterization of a hierarchy, as two topologically distinct hierarchies can be associated with the same $n^*$\cite{SAT21}. For instance, the C-cell hierarchy $(T_R C_L)$ and the E-cell hierarchy $(T_L E_L)$ both correspond to $n^* = 2$. Clearly, these two hierarchies are topologically distinct due to their differing $\sigma_{\pm}$ values. This underscores the importance of $\sigma_{\pm}$, which are absent in the above renormalization group (RG) equations, as essential elements for labeling a butterfly. \\

We conclude this section by noting that the aforementioned RG trajectories also emerge from pure number theory in two different contexts. Firstly, they appear as a symmetry of the Farey tree, as described in Appendix (C). Secondly, they arise when the butterfly fractal is mapped to the Integral Apollonian gasket, as explained in the next section.

\section{ Butterfly Apollonian Connection ($\bf {\mathcal {ABC}}$ ) }

\begin{figure}[htbp] 
 \includegraphics[width = .55 \linewidth,height=.55 \linewidth]{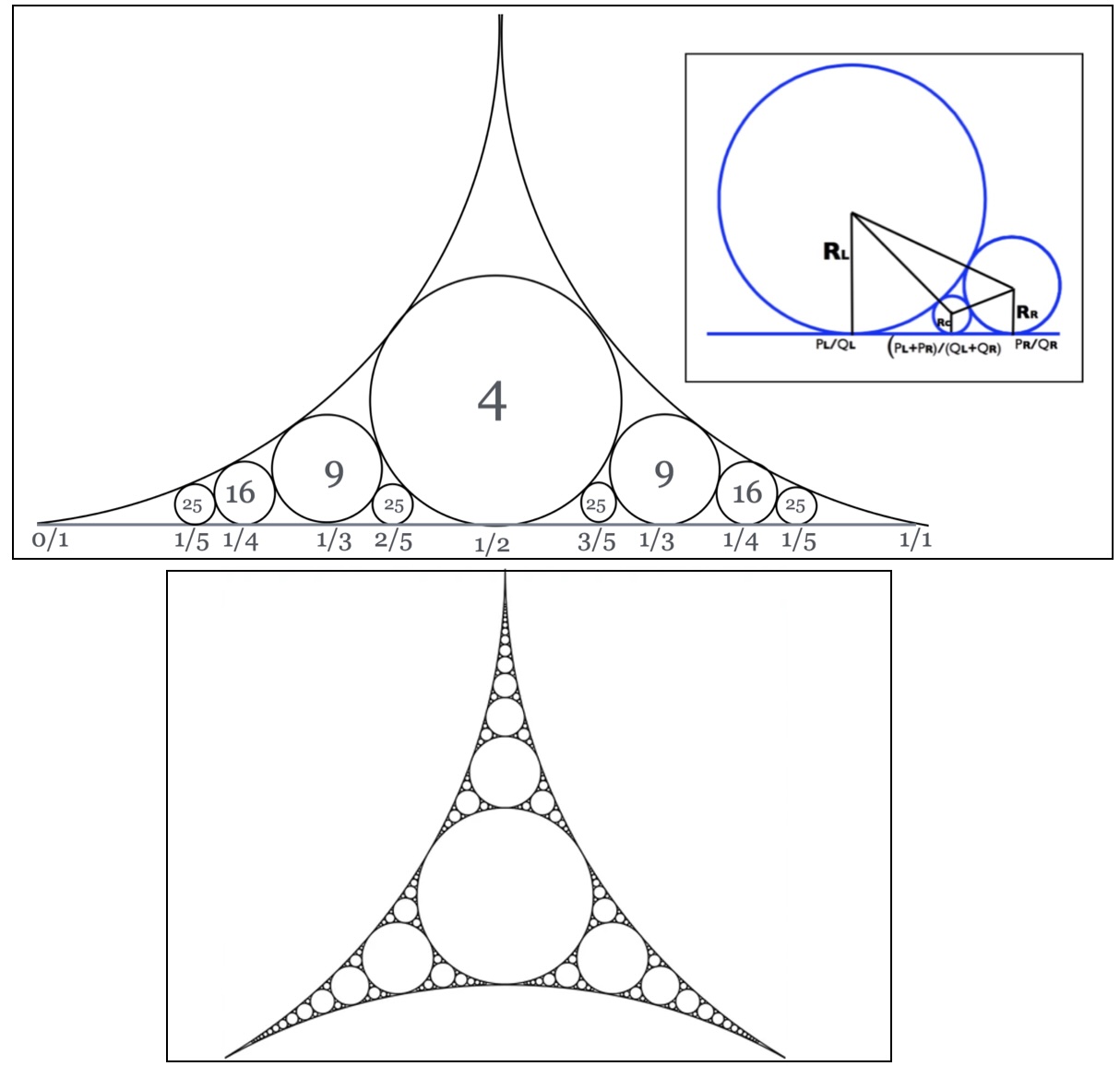} 
 \caption{ ( upper left ) Shows the Ford circles representing some fraction $\frac{p}{q}$. Curvatures of the circle shown with an integer inside the circle are scaled by a factor of $2$. The configuration of three blue circles on the right illustrate the fact that three rationals obeying Farey sum rule are represented by three mutually tangent  Ford circles. We will refer this gasket as ``Ford Apollonian" and is a subset of a general Apollonian shown below which is a configuration of four circles of non-zero curvatures. }
\label{bap}
\end{figure}

The story of ${\mathcal ABC}$—an intriguing relationship between the butterfly and the integral Apollonian gasket—stems from a remarkable discovery by American mathematician Lester Ford in 1938. Ford uncovered a pictorial representation of rational numbers using circles. He demonstrated that at each rational point $\frac{p}{q}$, where $p$ and $q$ are relatively prime, one can draw a circle with a radius of $\frac{1}{2q2}$ that is tangent to the x-axis in the upper half of the xy-plane. The curvatures of these Ford circles are $2q2$. A key characteristic of Ford circles is that two circles representing distinct fractions never intersect; they can only be tangent to each other. This tangency occurs when the two fractions are "Farey neighbors," or neighboring fractions in the Farey tree. Three mutually tangent Ford circles representing three fractions—$\frac{p_L}{q_L}$, $\frac{p_c}{q_c}$, and $\frac{p_R}{q_R}$—are all tangent to the x-axis and obey the Farey relation, as illustrated in Fig. (\ref{bap}). This relationship  among Ford circles is central to the concept of " ${\mathcal ABC}$ ," where three mutually tangent circles are associated with the butterfly triplet $[\frac{p_L}{q_L}, \frac{p_c}{q_c}, \frac{p_R}{q_R}]$.\\

Figure (\ref{bap}) depicts Ford circles forming a gasket, referred to as the "Ford Apollonian." This formation is a subset of the general Apollonian gasket—a close packing configuration of four circles, where three are tangent to each other, as shown in the lower panel. Before delving into ${\mathcal ABC}$, we briefly review the subject of the Apollonian gasket, named after studies from around 200 BC by Apollonius of Perga. Figure shows Ford circles forming a gasket- will  be referred as the ``Ford Apollonian". It is a subset of a general Apollonian gasket - a close packing of configurations of four circles where three are tangent to each other as shown in the lower panel. Before discussing ${\mathcal ABC}$, below is a  briefly review of the subject of Apollonian gasket, named after $200$ BC old studies by Apollonian of Perga\cite{book}.\\

Descartes' theorem states that the curvatures of the four circles in an Apollonian gasket satisfy the following equation:
\begin{equation}
 2( \kappa_1^2+\kappa_2^2+\kappa_3^2+\kappa_4^2) =  ( \kappa_1+\kappa_2+\kappa_3+\kappa_4)^2\\
 \label{Q4}
 \end{equation}

Although numerous proofs of this theorem exist \cite{Kocik}, its form suggests a simple and elegant proof that remains elusive. Appendix D presents a beautiful proof of Descartes' theorem by Richard Friedberg \cite{FR}, which applies to Ford circles (where $\kappa_4=0$). In this case, the theorem is demonstrated to be a consequence of a property of homogeneous functions in three variables.\\

Interestingly, Equation (\ref{Q4}) reveals how an Apollonian configuration can form a gasket where all circles have integer curvatures. The quadratic relationship among the four curvatures implies that, given any three mutually tangent circles with curvatures $(\kappa_1, \kappa_2, \kappa_3)$, there are precisely two possible circles that can be tangent to these three. In essence, given three mutually tangent circles, there are two distinct ways to complete a configuration of four mutually tangent circles, where the curvature of the fourth circle is determined by:
\begin{equation}
\kappa_{\pm}=(\kappa_1+\kappa_2+\kappa_3) \pm \sqrt{ \kappa_1 \kappa_2+\kappa_2 \kappa_3+ \kappa_3 \kappa_2}.
\end{equation}

 Denoting these two solutions as   $\kappa_4, \kappa'_4$, they satisfy a linear equation,
  \begin{equation}
 \kappa_4 + \kappa'_4 = 2 ( \kappa_1+\kappa_2+\kappa_3) 
 \label{l4}
 \end{equation}
Starting with three mutually tangent circles, we can therefore construct two distinct quadruplets $(\kappa_1, \kappa_2,\kappa_3, \kappa_4) $ and $(\kappa_1, \kappa_2,\kappa_3, \kappa^{\prime}_4)$.  A remarkable feature of the linear equation  (\ref{l4}) is that  if the original four circles have integer curvature, all of the circles in the packing will have integer curvatures. Furthermore, 
it leads to four generators that produce recursive scheme to generate the entire gasket. Forming the Apollonian group,
  the recursive filling of the space with Descartes configurations can be studied , adding additional circles  is accomplished by applying  four matrices $S_i,  ( i = 1-4)$ to a 
a root  configuration where the four matrices are:
\begin{eqnarray*}
S_1 =   \left( \begin{array}{cccc} -1 & 2 & 2  & 2 \\    0 & 1 & 0 & 0  \\  0 & 0 & 1 & 0\\ 0 & 0 & 0& 1  \\ \end{array}\right)\,\
S_2= \left( \begin{array}{cccc} -1 & 2 & 2  & 2 \\    0 & 1 & 0 & 0  \\  0 & 0 & 1 & 0\\ 0 & 0 & 0& 1  \\ \end{array}\right) \,\
S_3  =    \left( \begin{array}{cccc} -1 & 2 & 2  & 2 \\    0 & 1 & 0 & 0  \\  0 & 0 & 1 & 0\\ 0 & 0 & 0& 1  \\ \end{array}\right)\,\ S_4=
 \left( \begin{array}{cccc} -1 & 2 & 2  & 2 \\    0 & 1 & 0 & 0  \\  0 & 0 & 1 & 0\\ 0 & 0 & 0& 1  \\ \end{array}\right) \\
 \label{S4}
\end{eqnarray*}

\begin{figure}[htbp] 
 \includegraphics[width = .65 \linewidth,height=.7 \linewidth]{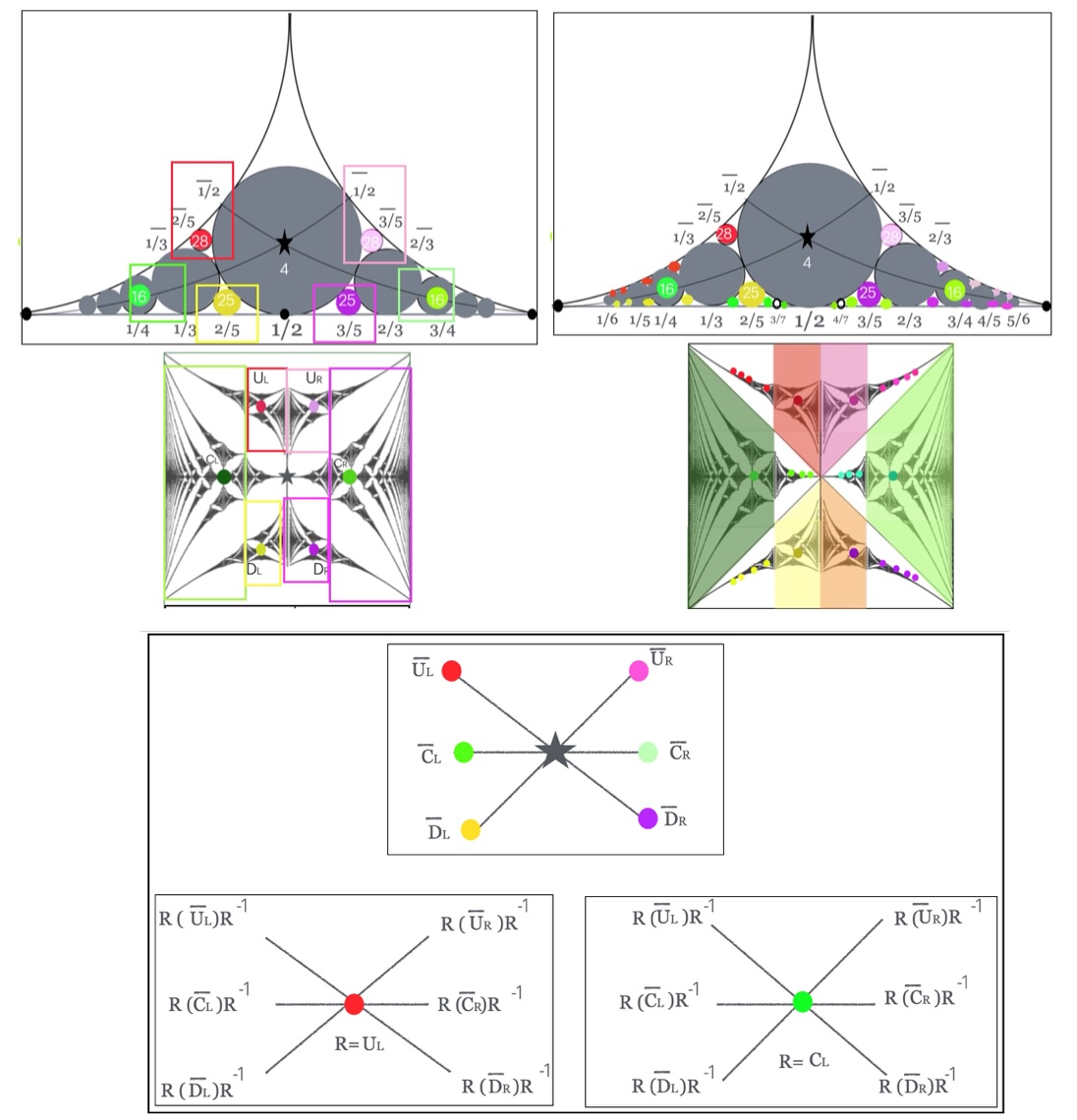} 
 \caption{The figure illustrates ${\mathcal ABC}$, showcasing the parent butterfly along with its six baby butterflies. Just as the "X"-shaped channel of gaps divides the butterfly into C-cell and E-cell butterflies, Pappus mirrors—circles passing through the tangency points of the Pappus chain (represented by chain of grey circles)—help compartmentalize the gasket into six regions that correspond to $(E_L, D_L, E_R, D_R, C_L, C_R)$. The centers of the parent butterflies are marked with a star, while the centers of the six baby butterflies are color-coded. The rectangles surrounding the centers highlight the portions of the Apollonian gasket associated with each butterfly.  The lowest panel shows the Apollonian tree that mimics the butterfly tree, highlighting the fact that to describe Apollonian recursions, every Apollonian configuration has to be first mapped to the Ford Apollonian. We note that, within the Apollonian configuration, all non-central tails are degenerate. }
\label{ba}
\end{figure}

Figure (\ref{ba}) illustrates ${\mathcal ABC}$, where butterfly centers are color-coded to correspond with the circles in the Apollonian gasket, representing the butterfly center. The rectangles, also color-coded, enclose segments of the four circles forming an Apollonian configuration associated with the butterfly. The main butterfly, with flux values at its boundaries and center at $[\frac{0}{1}, \frac{1}{2}, \frac{1}{1}]$, is depicted by a configuration of three mutually tangent Ford circles at $[\frac{0}{1}, \frac{1}{2}, \frac{1}{1}]$.\\

To describe the next generation of butterflies, we observe that given two tangent circles with curvatures $\kappa_1$ and $\kappa_2$, there exists a chain of circles where each pair of consecutive circles is tangent to one another. Known as Pappus chains, these configurations were first investigated by Pappus of Alexandria in the 3rd century. The tangency points of the chain lie on what is referred to as the "Pappus mirror," as it reflects $\kappa_1$ into $\kappa_2$ and similarly transforms the entire hierarchical set of circles tangent to $\kappa_1$ into those tangent to $\kappa_2$. In other words, for a hierarchical configuration where all circles share a common circle, termed the "boundary circle," there exists a twin configuration—its mirror image. Figure (\ref{ba}) illustrates two sets of chains, depicted with grey circles and black curved lines as Pappus mirrors, which reflect the x-axis to $\kappa_{\frac{0}{1}}$ and $\kappa_{\frac{1}{1}}$.\\

Pappus mirrors divide the gasket into distinct regions that host the E-cell butterflies and the C-cell representations, as illustrated in the figure. The centers of the C-cell butterflies are represented by circles along the Pappus chains, while the centers of the E-cells are positioned on the upper and lower boundary curves. Notably, the Apollonian representations of $E_L, D_L$ and $(E_R, D_R)$ are always mirror images of each other. Interestingly, although the E-cell butterflies lose mirror symmetry in the ($E-\phi)$ during the next generation, their Apollonian representation consistently retains this symmetry about the Pappus mirror.\\

The recursive structure of the Apollonian gasket, as described below, elucidates the explicit correspondence between the butterfly and the Apollonian. The theoretical framework employs M\"{o}bius transformations, which are briefly reviewed in Appendix D.  Basic idea is that since each Apollonian  is uniquely determined by three points \cite{Sat21IAG} and M\"{o}bius transformations relate two sets of three points, these transformations are utilized to transform one Apollonian to another. In establishing ${\mathcal ABC}$, these transformations are used in in two distinct ways:\\

(I)  Mapping the $x$-axis or Ford circles to an arbitrary circles through M\"{o}bius transformations belonging to the group $SL(2,C)$. For instance, the transformation that maps the x-axis to $\kappa_{\frac{0}{1}}$ is given by:
\begin{equation}
f(z) = \frac{z}{-i z+1}
  \label{cmap}
  \end{equation}

This transformation determines the mirror images of fractions $\frac{p}{q}$, depicted as $\bar{\frac{p}{q}}$ in the figure.\\

(II)  Two Ford Apollonians can be mapped to each other using M\"{o}bius transformations that form the $SL(2,Z)$ group. If the Apollonians are not initially Ford Apollonians, they are first mapped to Ford Apollonians using $SL(2,C)$. This process enables the mapping of any Apollonian within the gasket to another, facilitating the transformation of a parent Apollonian into six baby Apollonians.\\

To establish recursion relations associated with the nesting of the Ford circles, consider  two  sets of  triplets,
 $(z_1, z_2, z_3)$ and $(w_1, w_2, w_3)$ representing three rationals on the x-axis. which we will denote  as $[\phi_L(l), \phi_c(l), \phi_R(l)]$
and $[\phi_L(l+1), \phi_c(l+1), \phi_R(l+1)]$.
The recursion relation underlying the self-similar pattern is a fixed M\"{o}bius transformation  determined by mapping the triplets
  $( z_1, z_2, z_3)$ to the triple
 $( w_1,w_2, w_3) = (\frac{p^*_L}{q^*_L}, \frac{p^*_c}{q^*_c},\frac{p^*_R}{q^*_R}) =( \phi^*_L(1), \phi^*_C(1), \phi^*_R(1) )$.  We now illustrate this when the parent 
 is the Ford Apollonian representing three fractions $( 0, \frac{1}{2}, 1)$ which in turn represents the main butterfly. With $( z_1, z_2, z_3)= 0, \frac{1}{2}, 1)$,
 the constants $(a,b,c,d)$  of the map $w = f(z) = \frac{az+b}{cz+d}$ are determined as:
 \begin{equation}
 f(0)= \frac{b}{d}=\frac{p^*_L}{q^*_L},\,\,\,\,\ f(1/2)= \frac{a+2b}{c+2d}=\frac{p^*_c}{q^*_c},\,\,\,\,\ f(1)=\frac{a+b}{c+d}=\frac{p^*_R}{q^*_R}
 \end{equation}
 Using the  Farey relation (Eq. (\ref{FR})), we get
 \begin{equation}
 a=p^*_R-p^*_L,  \,\ b=p^*_L, \,\ c = q^*_R-q^*_L  ,  \, \,\ d = q^*_L
 \label{abcd1}
 \end{equation} 
 
 The corresponding $\phi= \frac{p}{q}$ recursions or equivalently $(p_x, q_x)$ recursions can be written as,
\begin{equation}
 \phi(l+1) =  \frac{ (p^*_R-p^*_L) \phi (l) + p^*_L}{ (q^*_R-q^*_L) \phi (l) + q^*_L} , \quad \left(\begin{array}{c}
p_x(l+1)\\
q_x(l+1)
\end{array}\right)   = \left( \begin{array}{cc}  p^*_R-p^*_L & p^*_L   \\  q^*_R-q^*_L &  q^*_L \\ 
\end{array}\right)
\left(\begin{array}{c}
p_x(l)\\
q_x(l)
\end{array}\right).
 \label{Mmap}
 \end{equation}

which is the same equation ( Eq. (\ref{R1}) )  that was derived using RG. Finally, when the mapping involves a non-Ford Apollonian—meaning the parent or its baby configuration is not a Ford Apollonian—the non-Ford Apollonian is first mapped to a corresponding Ford Apollonian. To illustrate this, consider the mapping corresponding to $U_L$, where $(z_1, z_2, z_3) = (0, \frac{1}{2}, 1)$ is mapped to $(w_1, w_2, w_3) = (\bar{\frac{1}{3}}, \bar{\frac{2}{5}}, \bar{\frac{1}{2}}) = (\frac{3+i}{10}, \frac{10+4i}{29}, \frac{2+i}{5})$. In such cases, the non-Ford Apollonian $(w_1, w_2, w_3)$ is mapped to a Ford Apollonian using Equation (\ref{cmap}). Subsequently, Equation (\ref{Mmap}) is used to determine the transformation. By following this procedure for the six baby configurations shown in Figure (\ref{ba}), we obtain six generators:
\begin{eqnarray}
\label{EE1}
\bar{U}_L  & =  & \left( \begin{array}{cc} i  &  1 \\ 3i & 3-i \\ 
\end{array}\right),\,\,\ \bar{D}_L   =   \left( \begin{array}{cc} 0  &  1 \\ -1 & 3 \\ 
\end{array}\right),\,\,\  \bar{C}_L   =  \left( \begin{array}{cc} 1  &  0 \\ 2 & 1 \\ 
\end{array}\right)\\
\bar{U}_R  & = &  \left( \begin{array}{cc} 0  &  1 \\ -1 & 3-i \\ 
\end{array}\right), \,\,\ \bar{D}_R = \left( \begin{array}{cc} 1  &  1 \\ 1 & 2 \\  
\end{array}\right), \,\,\ \bar{C}_R = \left( \begin{array}{cc} -1  &  2 \\ -2 & 3 \\   
\end{array}\right)
\label{EE2}
\end{eqnarray}
The eigenvalues of $(\bar{U}_L,\bar{D}_L)$ ($(\bar{U}_R,\bar{D}_R)$) are $\frac{3\pm\sqrt{5}}{2}$ and those of $(\bar{C}_L,\bar{C}_R)$ are unity.\\
The above six generators describing Apollonian nesting are elements of $(SL(2,\mathbb{Z}[i])$ where $\mathbb{Z}[i])$ represents the Gaussian integers - complex numbers whose real and imaginary parts are integers. They correspond  to the six generators $(U_L, D_L, C_L, U_R, D_R, C_R)$ in Equations (\ref{E1}-\ref{E2}) that characterize the butterfly fractal.  It may appear that the generators $(\bar{U}_L,\bar{D}_L, \bar{U}_R,\bar{D}_R,\bar{C}_L,\bar{C_R}) $  actually correspond to the generators $(E_L, E_R, C_L, C_R)$ of Eq. (\ref{M2}) as these two sets same eigenvalues: both have unit determinant and trace 3 and trace 2 for edge and center configurations respectively. However $E_x$ generates both the $U_x$ and $D_x$ butterflies -  butterflies with same  magnetic flux interval but have different topological characteristics.
This shortcoming was remedied by introducing $ 3 \times 3 $ matrices as shown in Eqs (\ref{T1}-\ref{T2}). The generators  in equations (\ref{E1}-\ref{E2}) distinguishes $U_x$ and $D_x$
as these two generators have different eigenvectors. It is conceivable that $\sigma_{\pm}$ although invisible in the $\mathcal{ABC}$,  are hidden in the eigenfunctions.
Finally, we note that within $\mathcal{ABC}$, non-central tails are degenerate - a shortcoming that may be remedied by using Ford spheres- a generalization that remains an open problem.\\

\section{ Butterfly Fractal and Tree of  Pythagorean Triplets }
Readers captivated by the union of the butterfly and Apollonian are in for an even greater surprise as they uncover mathematics dating back to 500 BC embedded within this fractal. Remarkably, the recursive structure of the C-cell butterflies and the butterfly tails is intricately linked to the tree of primitive Pythagorean triplets \cite{SAT21,SAT18}. These parts of this fractal are 
described by three integers satisfying the equation $(a^2 + b^2 = c^2)$. The method for constructing this butterfly is subtly hidden within the tree of Pythagorean triplets, known for almost a century, which recursively generates the entire set of triplets.\\

\begin{figure}[htbp] 
\includegraphics[width = .65\linewidth,height=0.6\linewidth]{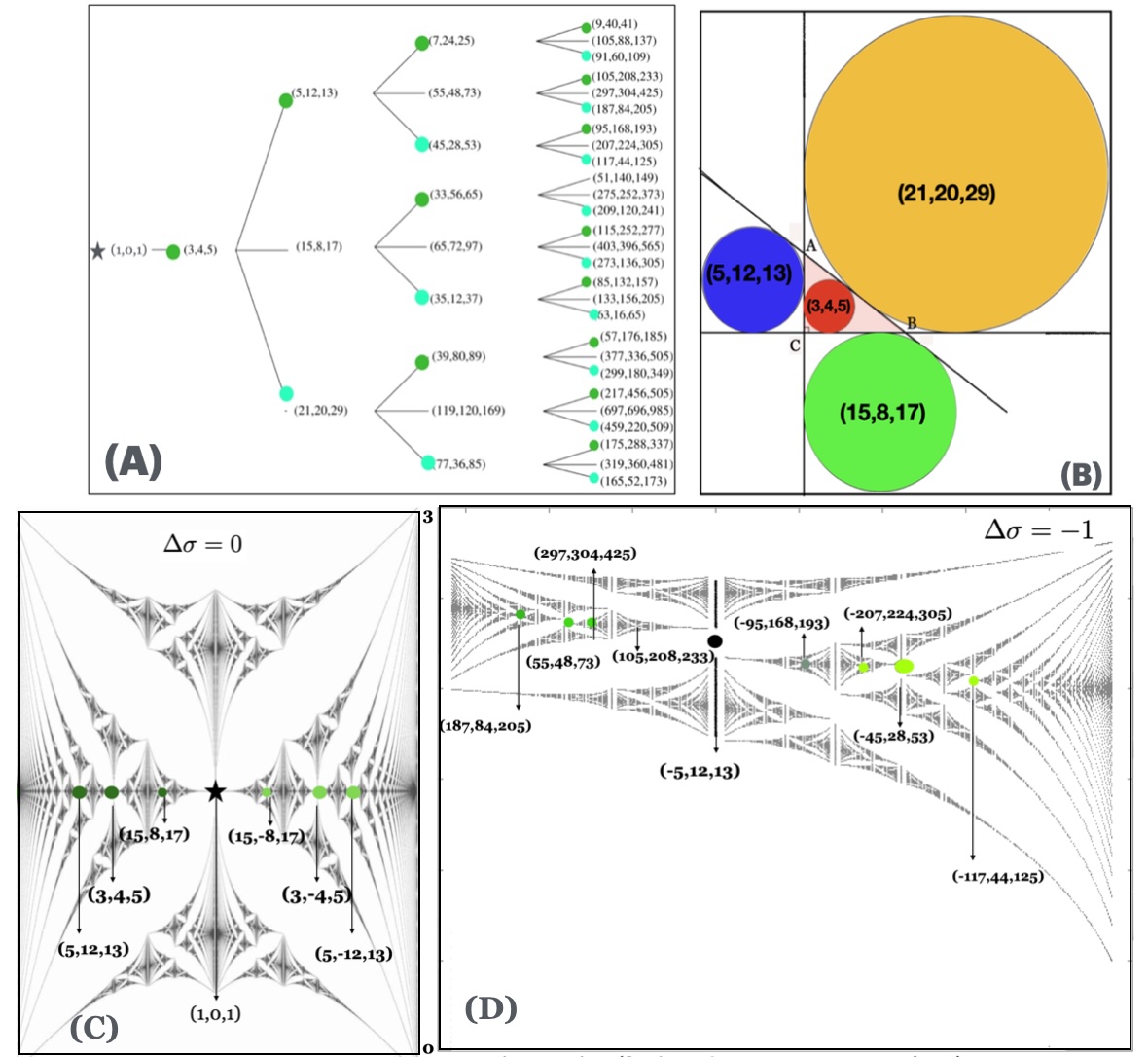}
\leavevmode \caption{ Panel (A) depicts the tree of all Pythagorean triplets. The ternary nature of the tree \cite{c3} is demonstrated in panel (B), where the fundamental concept is that for any right-angled triangle ABC, there is an incircle and three excircles \cite{c3}, with the excircles becoming incircles at the next level. Panels (C-D) showcase butterflies labeled with Pythagorean triplets, where some entries include negative members in the triplet. Additionally, the tree associated with these butterflies features an extra label, $\Delta \sigma$.}
\label{PT}
\end{figure}

Given the butterfly integers $(q_R, q_L)$, we can construct primitive Pythagorean triplets $(a,b,c)$ where $a^2+b^2=c^2$ as:
 \begin{eqnarray}
\label{ep1}
(a,b,c)  & =   & \Big( q_L q_R , \frac{q_R^2-q_L^2}{2}, \frac{q_R^2+q_L^2}{2} \Big),  \rm{   q_R, q_L \,\ have \,\  same \,\ parity }\\
& =   & \Big (q_R^2-q_L^2, 2q_R q_L, q_R^2+q_L^2 \Big),  \rm{  q_R, q_L \,\ have \,\  opposite \,\ parity } \nonumber
\label{ep2}
\end{eqnarray} 

It has been known 
that  the entire set of all primitive Pythagorean triples has the structure of a ternary tree\cite{Hall} (see Fig. (\ref{PT})).
The tree is generated iteratively by three matrices $H_1$, $H_2$, $H_3$, acting on the column matrix with entries as the Pythagorean triplet $(a,b,c)$:
\begin{equation}
H_1= \left( \begin{array}{ccc} 1 & -2 & 2  \\    2 & -1 & 2  \\  2 & -2  & 3 \\ \end{array}\right),
H_2=\left( \begin{array}{ccc} 1 & 2 & 2  \\    2 & 1 & 2  \\  2 & 2  & 3 \\ \end{array}\right),
H_3=\left( \begin{array}{ccc} -1 & 2 & 2  \\    -2 & 1 & 2  \\  -2 & 2  & 3 \\ \end{array}\right) 
\label{H3}
\end{equation}

One can also describe the tree  using
 $ 2\times 2$ matrices, acting on the pair  $(q_R,q_L)$:
\begin{equation}
h_1 =  \left( \begin{array}{cc} 1   & 2  \\  0 & 1   \\  \end{array}\right), \quad  h_2 =  \left( \begin{array}{cc} 2   & 1  \\  1 & 0   \\  \end{array}\right),\,\,\,\ h_3 =  \left( \begin{array}{cc} 2   & -1  \\  1 & 0   \\  \end{array}\right)
\end{equation}

To connect the butterfly recursions with the Pythagorean tree, we must incorporate two additional features into the tree structure. Firstly, it is necessary to include negative entries in the tree as $b$ can take both positive and negative values ( see Eq. (\ref{ep2}) . To accommodate these two possibilities, we need to introduce four matrices. It is advantageous to work with  $2 \times 2$ matrices, as they generate new sets of Euclidean parameters and therefore relate more directly to the butterfly labeling. By relabeling the matrices to represent the left and right offspring of the parent butterfly, the four matrices are:

\begin{equation}
c_L =  \left( \begin{array}{cc} 1   & 2  \\  0 & 1   \\  \end{array}\right),\,\ c_R =  \left( \begin{array}{cc} 1   & 0  \\  2 & 1   \\  \end{array}\right), \,\  \tau_{R}=  \left( \begin{array}{cc} 2   & -1  \\  1 & 0   \\  \end{array}\right), \,\  \tau_{L} =\tau^{-1}_{ R}, 
\end{equation}

Secondly, the Pythagorean tree necessitates a new label: $\Delta \sigma$, as depicted in the figure. Consequently, within the central region of every sub-butterfly lies the Pythagorean tree, with each tree possessing an integer designation that encodes the symmetry of the butterfly.\\

We will now summarize the  relationship between the three generators $(h_1, h_2, h_3)$ of the Pythagorean tree,  the four generators $(S_1, S_2, S_3, S_4)$ and their mirror  images  $(\bar{S_1}, \bar{S_2}, \bar{S_3}, \bar{S_4})$
of the Apollonian gasket and the eight generators  of the butterfly tree.
\begin{eqnarray*}
h_1 \equiv c_L  & \leftrightarrow &   S_1 S_2, \,\,\ h_2  \equiv c_R \leftrightarrow S_1 S_3 ,\,\,\  h_3 \equiv \tau_L \leftrightarrow  S_1 S_3 S_1, \,\,\ \tau_R = (S_1 S_3 S_1)^{-1}\\
\\
 U_L &  \rightarrow  &  S_4 S_2 ,\,\,\   U_R \rightarrow  S_4 S_3 ,\,\,\ D_L  \rightarrow \overline{S_4 S_2} , \,\,\ D_R \rightarrow  \overline{S_4 S_3}.
\label{HS}
\end{eqnarray*}

We close this section by pointing out that although the Pythagorean tree describes only a subset of the butterfly ( C-cell only), the butterfly tree is actually a tree of quadruplets\cite{Sat21IAG} - 
 $( a,b,c,d)$ , known as
the  Lorentz quadruplets,  where
 $ a^2+b^2+c^2=d^2$. This is due to the following relationship between 
the four curvatures  $(\kappa_1, \kappa_2, \kappa_3, \kappa_4)$ of any Apollonian configuration  and the quadruplets:
 
\begin{eqnarray}
 \left( \begin{array}{c} a  \\    b  \\  c  \\  d  \\ \end{array}\right) =  \left( \begin{array}{cccc} 1 & -1 & -1  & -1 \\    0 & 0 & 0 & 2  \\  0 & 1 & -1 & 0\\ 1 & 1 & 2& 1  \\ \end{array}\right) 
  \left( \begin{array}{c} \kappa_1  \\  \kappa_2  \\  \kappa_3 \\  \kappa_4 \\ \end{array}\right) . \end{eqnarray} 
  
  However, it appears that the Lorentz quadruplets associated with the four curvatures do not necessarily have to be primitive, as observed in the case of the golden hierarchy. Consequently, the Apollonian tree—a counterpart to the butterfly tree—essentially comprises quadruplets that include non-primitive entries. To the best of our knowledge, there is no existing quadruplet tree capable of generating all primitive or non-primitive Lorentz quadruplets.
 \section{ Butterfly meets Mandelbrot  }
  \begin{figure}[htbp] 
 \includegraphics[width = .65 \linewidth,height=.5 \linewidth]{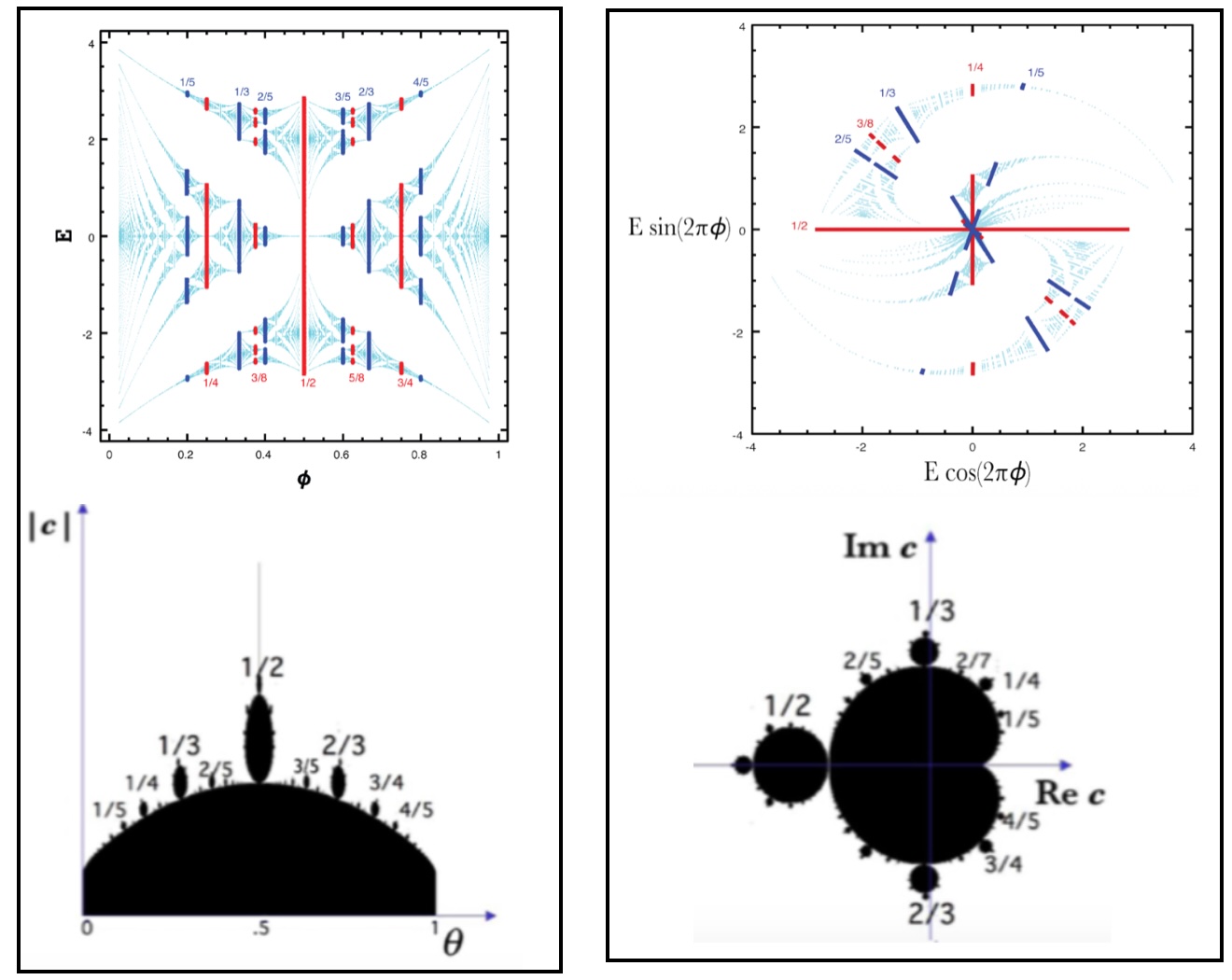} 
 \caption{ Each of the two panels showcases the butterfly fractal and the Mandelbrot Set in distinct ways. Both are two-dimensional fractals, where the variable (z) in the Mandelbrot Set corresponds to $Ee^{i 2\pi \phi}$ in the butterfly fractal. The bands in the butterfly and the bulbs tangent to the cardioid in the Mandelbrot Set are labeled with a rational number $\frac{p}{q}$, revealing the Farey tree with the Farey sum rule embedded within these two fractals.}
\label{MS}
\end{figure}
It is truly a marvel of nature that two fractals—the Butterfly Fractal and the Mandelbrot Set—which emerged around the same time, share a remarkable common feature. The Farey sum rule, which forms the backbone of the butterfly fractal, is also a significant aspect of the Mandelbrot Set\cite{book}:
\begin{equation}
z_{n+1} = z_n +c ,\,\ ,\ z_0 = 0
\end{equation}
 To make this correspondence explicit, we write Harper's equation  ( Eq. (\ref{harper}) as a dynamical map:
\begin{eqnarray*}
r_{n+1} & =  & \frac{1}{r_n+E + 2 \cos( 2 \pi \theta_n)}\\
\theta_{n+1} & = & \theta_n + 1\,\  {\rm mod} \,\ 1,
\label{rt}
\end{eqnarray*}
where $r_n= \frac{ \psi_n}{ \psi_{n-1}} $ and $\theta_n = 2 \pi n \pi + k_y$. The butterfly graph is obtained by using the set $(\phi, E)$ to satisfy the following equation:

\begin{equation}
\gamma =  \lim_{ N \rightarrow \infty} \frac{1}{2N} log ( \frac{r_1}{r_N} )=0
\end{equation}

The above equation  simply means that the butterfly graph consists of values of $\phi$ and $E$
for which the two-dimensional mapping given above does not diverge,
which is to say, it has a Lyapunov exponent of $0$. The
negative Lyapunov exponents give the gaps of the butterfly.\\

To underscore the commonality between the butterfly fractal and the Mandelbrot Set, figure (\ref{MS}) showcases two distinct representations of the butterfly fractal alongside two depictions of the Mandelbrot Set. In the butterfly fractal, the variables $(r, \theta)$ specified in equation (\ref{rt}) correspond to the variables encoded within the complex variable $z$ of the Mandelbrot Set. Similarly, the parameters $(E, \phi)$ play the role of the complex parameter $c$ of the Mandelbrot Set. The condition of a zero Lyapunov exponent in the Hofstadter set mirrors the requirement that the permissible values of $c$ correspond to a bounded set of values for $z$. This comparison is vividly illustrated in the figure. An intriguing shared feature of these two sets is the emergence of regions—bulbs in the Mandelbrot Set and bands in the butterfly—characterized by simple rational fractions with striking commonality being the Farey organization of rationals. Thus, number theory plays a pivotal role in both the Mandelbrot Set and the butterfly fractal.

 \section{ Seeing Butterfly fractal in Laboratories}
 Hofstadter's butterfly was initially predicted under experimental conditions that were once unattainable using laboratory-scale magnetic fields. However, recent electrical transport studies have provided evidence for Hofstadter's butterfly in materials engineered to have artificially large lattice constants, such as those with moiré superlattices\cite{Dean, Nature2025}. These superlattices are created by stacking and twisting two sheets of carbon atoms, forming an electron pattern reminiscent of the moiré design, a common French textile, made possible through breakthroughs in materials engineering. The mathematical intrigue of this problem arises from the fact that a bilayer forms a two-dimensional crystal only at a discrete set of commensurate rotation angles. Theoretical models\cite{MV} predict multiple "magic angles" where tunneling becomes maximal, with the largest of these angles being the most feasible to achieve experimentally.\\

A team of scientists from Princeton used a scanning tunneling microscope (STM) to image moiré crystals at atomic resolution and examine their electron energy levels. The STM operates by bringing a sharp metallic tip less than a nanometer from the surface, allowing quantum "tunneling" of electrons from the tip to the sample. This tool was crucial for the experiment, as it is particularly sensitive to the energy of electrons in materials, serving as a direct energy probe that correlates with Hofstadter's original calculations of energy levels. Previous studies on Hofstadter's butterfly relied on electrical resistance measurements, which do not directly measure energy. The intrigue lies in the comparable scales of MATBG moiré and magnetic lengths for experimentally accessible magnetic fields. Focusing on the insulating states, a rich phase diagram in terms of magnetic flux versus the filling factor has been observed. For magnetic fields around 27 Tesla, an entire flux quantum per moiré unit cell is achieved, allowing the observation of physics related to Hofstadter's fractal structure of electronic gaps.\\

The study of the electronic properties of magic angle twisted bilayer graphene (MATBG) in a magnetic field is an active area of research. 
Interestingly, the researchers were not initially searching for the butterfly; they were investigating superconductivity in twisted bilayer graphene. The studies conducted by Yazdani's group mark the first time that fractality in the energy of electrons in a two-dimensional crystal under a magnetic field has been directly observed in a real material.  Furthermore, their studies were able to see the effects of  electron-electron interactions on the spectrum,  an important feature that was not explored by Hofstadter's original calculations.
\section{ Closing}

The Hofstadter Butterfly remains a landmark discovery, and the allure of this quantum fractal continues to grow as its fractality, once confined to theoretical predictions, has now been observed experimentally \cite{Dean, Nature2025}. A distinctive feature of the butterfly fractal is its composition of integers—the topological quantum numbers of Hall conductivity. It is truly remarkable to realize that, conceived in 1976, the butterfly fractal encodes mathematical concepts discovered as early as 300 BC and incorporates the quanta of the quantum Hall effect discovered in 1983. In a 1939 lecture, Paul Dirac noted that "pure mathematics and physics are becoming ever more closely connected," suggesting a potential unification where "every branch of pure mathematics then having its physical application." Undoubtedly, the butterfly problem reveals profound connections between elegant mathematics and Nature's embrace of it in designing the world.\\

A treasure trove for physicists and mathematicians, the butterfly problem continues to offer delightful surprises. Its rich history includes connections to the Bethe Ansatz \cite{BA}, the prime number theorem \cite{prime}, and the Riemann hypothesis \cite{RH} through its link to the Apollonian gasket. A recent study titled “Spectral Theory of Extended Harper’s Model and a Question by Erdös and Szekeres” \cite{ES} hints at many more mathematical treasures hidden within this problem. Furthermore, exploring the butterfly Hamiltonian to investigate the effects of interactions\cite{Shan}, superconductivity \cite{HZS}, and new topological states of matter through time-dependent perturbations \cite{cp,SZ} is emerging as a vibrant area of research. This field is invigorated by experimentalists' fascination with bringing this butterfly and its variations to life in laboratories, including state-of-the-art experiments with ultra-cold atoms\cite{cold}. There is optimism that the interplay between theory and experiment may pave the way for the discovery of new classes of materials with technological payoffs, adding a new dimension to this problem.\\

As we progress, the Hofstadter Butterfly stands as a testament to the power of interdisciplinary science, demonstrating that the beauty of fractals transcends pure mathematics and touches the fundamental laws of the universe. Beneath its complexity lies simplicity. This also raises questions about the role of number theory—mathematics that often falls outside the vocabulary of most physicists—in the promising field of topological insulators. We close with this famous note by Galileo: "[Nature] is written in the language of mathematics, and its characters are triangles, circles, and other geometric figures without which it is humanly impossible to understand a single word of it; without these, one wanders about in a dark labyrinth."
\\
\\

\section{Acknowledgments}
My  sincere gratitude to Michael Wilkinson,  Richard Friedberg, and Jerzy Kocik for  many stimulating discussions and invaluable insights throughout this work.
On this $50$th anniversary of the Hofstadter butterfly, I want to express my deep appreciation to Doug Hofstadter and Francisco Claro. Their foundational work not only inspired me but also sparked an addiction to this problem—an endless chase that continues to this day.
\begin{figure}[htbp] 
 \includegraphics[width = .26 \linewidth,height=.37 \linewidth]{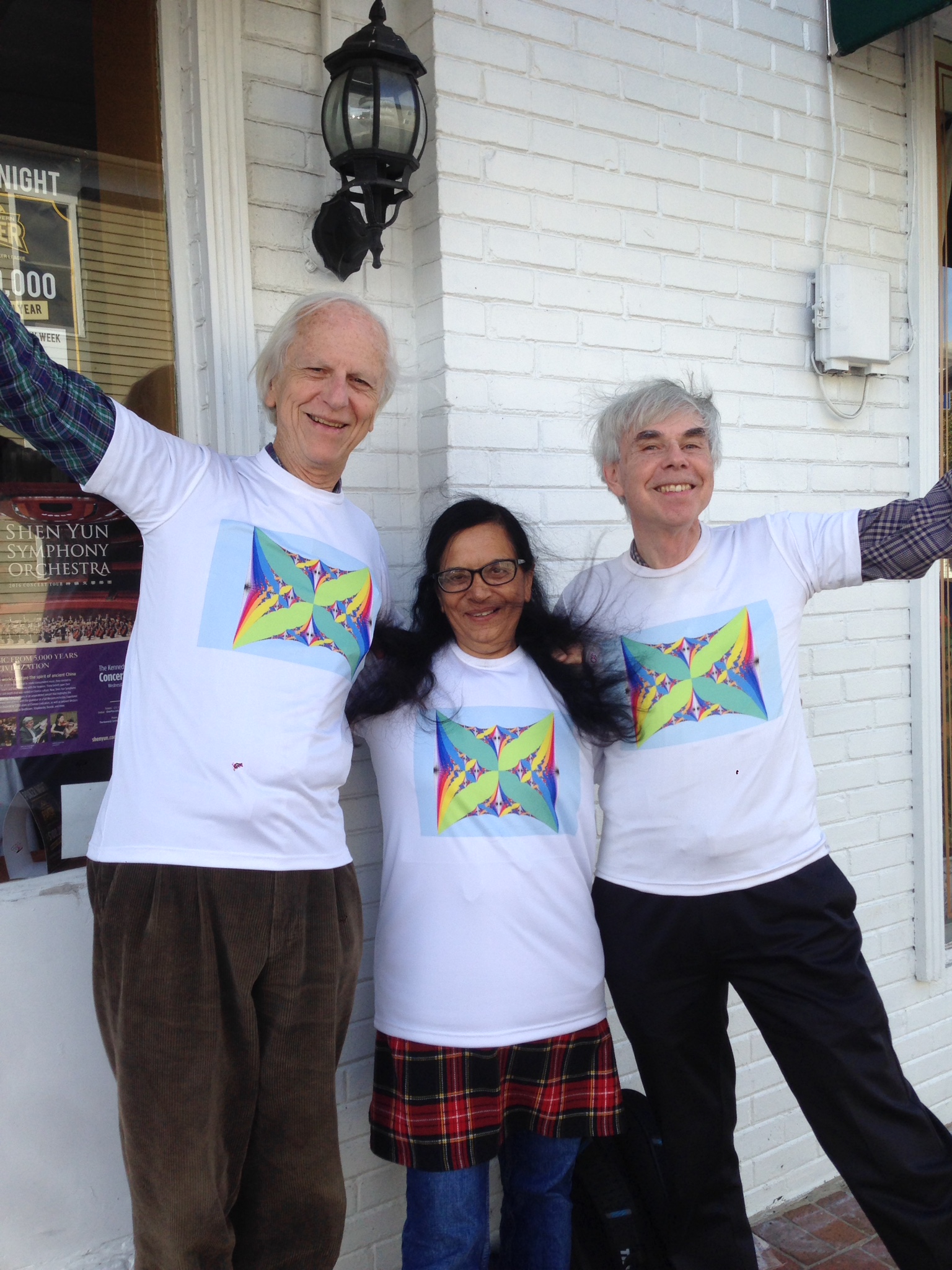} 
 \caption{ Author with Fransisco Claro and Doug Hofstadter in $2016$ }
\end{figure}

\appendix
\section{ Some Exact Results for the Spectrum }
 
  Coexisting simplicity and complexity is a norm in  the Hofstadter butterfly. One of the earliest evidence of this is in the Chamber's formula\cite{Chamber} which states that the energies $E$
 of the Harper's equation, for magnetic flux $\phi= p/q$  depend upon $k_x$ and $k_y$ -- the Bloch vectors in $x$ and $y$ direction, via $\Lambda$  which is equal to the determinant of $H$,
 
 \begin{eqnarray*}
 \Lambda ( k_x, k_y )  & = & 2 \cos ( q k_x) + 2 \cos( q k_y)\\
 &  = & E^q + a_1 E^{q-1} + a_2 E^{q-2}+.......
 \end{eqnarray*}
 Here the coefficients of the polynomial $a_i$ are independent of $k_x$ and $k_y$.
 Analytic expressions for the energy dispersions $E(k_x, k_y)$ for a few simple cases illustrate this intriguing result.
 \begin{itemize}

\item For $\phi = 1$, the energy spectrum consists of a single band given by $E =  2 ( \cos k_x + \cos k_y )$.

\item For $\phi=1/2$, the two bands having energies $E_+$ and $E_-$ are given by $E_{\pm}=\pm 2 \sqrt{ 1+ \frac{1}{2}( \cos2 k_x + \cos2 k_y)}$.

\item For $\phi=1/3$, the three bands have energies  $E_{n} =   2\sqrt{2} \cos ( \theta \pm  n \frac{2}{3} \pi)$. Here, $\theta= \frac{1}{3} Arccos[( \cos 3k_x+ \cos 3k_y)/2\sqrt{2}]$, $ n = 0,1,2$.

\item For $\phi = 1/4$, the energies of four bands are given by the expression $E= \pm \sqrt{4 \pm 2[3+\frac{1}{2} (\cos 4k_x + \cos 4 k_y)]^2}$.
\end{itemize}


 \section{ Proof of Farey Sum Rule }

The  empirical fact that every butterfly in the butterfly graph obey Farey sum rule ( Eq. (\ref{FR}) ) was put on a solid footing\cite{SW} using earlier renormalization studies 
of the Harper equation\cite{MW87} that proved the following equation:
\begin{equation}
\label{MW1}
\phi= \frac{p_0 + M_0 \phi'}{ q_0 - N_0 \phi'},
\end{equation} 
where $(M_0, N_0)$ are topological integers of one of the band at $\phi_0 = \frac{p_0}{q_0}$ that satisfy the equation

\begin{equation}
\label{eq: 2.1.2}
1=q_0M_0+p_0N_0,
\ .
\end{equation}
Here $N_0$ is the quantum number of Hall conductivity of the band. Here $\phi'$ is the renormalization of $\phi$ which is near $\phi_0$. We note that the renormalization describes
the behavior of just one band.\\

Applying this equation to a sub-butterfly described by a triplet $[\frac{p_L}{qL}, \frac{p_c}{q_c}, \frac{p_R}{q_R}]$,  set $\phi = \phi_0= \frac{p_L}{q_L}$ in Eq. (\ref{MW1}). This gives
$\phi'=0$ and therefore this $\phi$ value can be identified with the  flux value $\frac{p_L}{q_L}$ - the left boundary of the sub-butterfly. The right boundary with correspond to
$\phi = \frac{p_R}{q_R}$ with $\phi'=1$ and center of the butterfly with $\phi= \frac{p_c}{q_c}$ with $\phi'=\frac{1}{2}$. That is,
\begin{eqnarray*}
\frac{p_R}{q_R}  & = &\frac{p_L + M_0}{ q_L-N_0}\\
\frac{p_c}{q_c}  & = &\frac{p_c + M_0/2}{ q_L-N_0/2}
\end{eqnarray*}
The above two equations lead to the Farey sum rule: $ \frac{p_c}{q_c} = \frac{p_L+p_R}{q_L+q_R}$.

\subsection{ Farey sum rule gives butterfly recursion}

 Using the fact that every sub-butterfly follows the Farey Sum rule ( Eq. \ref{FR}), we will now derive the butterfly recursions using an 
 an  important symmetry property of the
Farey tree.   Here by the word  ``symmetry" , we do not refers to the Euclidean  geometrical symmetry, but symmetry described by
invertible algebraic transformations that maps one pair of Farey fractions to another.  That is, we seek a
transformation $T$ :
 
 \begin{equation}
  \Big(\frac{p_x(1)}{q_x(1)}, \frac{p_y(1)}{q_y(1)}\Big) \rightarrow  \Big(\frac{p_x(2)}{q_x(2)}, \frac{p_y(2)}{q_y(2)}\Big) = T  \Big(\frac{p_x(1)}{q_x(1)}, \frac{p_y(1)}{q_y(1)}\Big),
  \end{equation} 
 where each pair satisfies $ \Big(p_x(l)q_y(l)-p_y(l)q_x(l)\Big) =  1$)  and  the mapping preserves the order, that is $\frac{p_x(1)}{q_x(1)} \rightarrow \frac{p_x(2)}{q_x(2)}$ and $\frac{p_y(1)}{q_y(1)} \rightarrow \frac{p_y(2)}{q_y(2)}$. To obtain $T$, we construct  two matrices $T_1$ and $T_2$ as:
 \begin{equation}
T_1 =   \begin{bmatrix} p_L(1) & p_R(1) \\ \\ q_L(1) & q_R(1) \end{bmatrix},\,\,\,\ T_2=  \begin{bmatrix} p_L(2) & p_R(2) \\ \\ q_L(2) & q_R(2) \end{bmatrix}
\label{t12}
\end{equation}

We will now show that  required map is\cite{Hatcher}:
\begin{equation}
 T  = T_2 T^{-1}_1
  =    \frac{1}{D}  \begin{bmatrix} p_L(2) q_R(1)-p_R(2)q_L(1) & p_L(1)p_R(2)-p_L(2)p_R(1) \\  \\ q_L(2)q_R(1)-q_L(1)q_R(2) & p_L(1) q_R(2)-p_R(1)q_L(2)  \end{bmatrix} 
 \label{t}
 \end{equation}

To prove Eq. (\ref{t}), consider a transformation that maps  a primitive fraction $\frac{p}{q}$ to another primitive fraction $\frac{p^{\prime}}{q^{\prime}}$, defined as,
 
 \begin{equation}
 \frac{p}{q} \rightarrow \frac{p^{\prime}}{q^{\prime}} = \frac{a p+ b q}{c p + d q} \equiv \frac{ a\frac{p}{q} + b} { c \frac{p}{q} + d},
 \label{T}
 \end{equation}
 
 The above equation can also be written as,
\begin{equation}
  \left( \begin{array}{cc} p \\ q \end{array} \right) \rightarrow \left( \begin{array}{cc} p^{\prime} \\ q^{\prime} \end{array} \right) =  \left( \begin{array}{cc} a & b \\ c & d 
\end{array} \right) \left( \begin{array}{cc} p \\ q \end{array} \right) \equiv \mathcal{M}  \left( \begin{array}{cc} p \\ q \end{array} \right)
\end{equation}

Under this transformation, $\frac{0}{1} \rightarrow \frac{b}{d}$ and $\frac{1}{0} \rightarrow \frac{a}{c}$. In other words, $\mathcal{M}^{-1}$ maps a pair of fractions $(\frac{b}{d}, \frac{a}{c} )$ to  $( \frac{0}{1}, \frac{1}{1})$.\\

Therefore, the transformation that maps $ \Big(\frac{p_x(1)}{q_x(1)}, \frac{p_y(1)}{q_y(1)}\Big)$ to   $ \Big(\frac{p_x(2)}{q_x(2)}, \frac{p_y(2)}{q_y(2)} \Big)$ can be constructed as a two step process where we first
 map $ \Big(\frac{p_x(1)}{q_x(1)}, \frac{p_y(1)}{q_y(1)}\Big)$
 to $( \frac{0}{1}, \frac{1}{1})$  where $\left( \begin{array}{cc} a & b \\ c & d 
\end{array} \right) = T_1^{-1}$ and then map $( \frac{0}{1}, \frac{1}{1})$ to $ \Big(\frac{p_x(2)}{q_x(2)}, \frac{p_y(2)}{q_y(2)}\Big)$ where $\left( \begin{array}{cc} a & b \\ c & d 
\end{array} \right) = T_2$.

This completes the proof that $T = T_2 T_1^{-1}$ where $(T_1, T_2)$ are given by equation  (\ref{t12}).\\
\\

For self-similar hierarchical structure,  the renormalization equation connecting two consecutive levels $l$ and $l+1$ is given by,
\begin{equation}
  \left( \begin{array}{cc} p(l+1) \\ q(l+1) \end{array} \right) = T   \left( \begin{array}{cc} p(l) \\ q(l) \end{array} \right)
\end{equation}

\section{ M\"{o}bius Transformations}

Conformal maps are functions in complex plane that preserve the angles between curves.
There is a specific family of conformal maps  $w = \frac{az+b}{cz+d}$,  known as M\"{o}bius transformation or linear fractional transformation. Such  transformation maps lines and circles to lines and circles.
These maps can be represented  ( denoted by symbol $\dot{=}$ ) with a matrix:
 The map
\begin{equation}
 w = f(z)= \frac{az+b}{cz+d} \,\ \dot{=} \,\ \left( \begin{array}{cc} a & b \\ c & d \\
\end{array} \right) 
 \end{equation}

This identification of M\"{o}bius map with a matrix is useful because if we
consider two conformal maps: $ w_1= f_1(z)$ and $w_2=f_2(z)$ that respectively correspond to the matrices $C_1$ and $C_2$. Then the composition  $f_1f_2(z)$ corresponds to the matrix $C_1\cdot C_2$.

The  constants of Mobius maps $(a,b,c,d)$  can be determined in terms of 
two sets of triplets:  $( z_1, z_2, z_3) $ and   their conformal image $ ( w_1, w_2, w_3)$:

\begin{eqnarray}
 a & = & \det { \left( \begin{array}{ccc} z_{1} w_1 & w_1 &1\\ z_{2} w_2 & w_{2} &1\\  z_{3} w_{3} & w_{3} &1  \end{array}\right) } ,\,\
 b  =  \det {  \left( \begin{array}{ccc} z_{1}w_{1}&z_{1}&w_{1}\\z_{2}w_{2}&z_{2}&w_{2}\\z_{3}w_{3}&z_{3}&w_{3}  \end{array}\right) } \nonumber \\
  c & = & \det {  \left( \begin{array}{ccc} z_{1}&w_{1}&1\\z_{2}&w_{2}&1\\z_{3}&w_{3}&1  \end{array}\right)} ,\,\
 d    =   \det {  \left( \begin{array}{ccc} z_{1}w_{1}&z_{1}&1\\z_{2}w_{2}&z_{2}&1\\z_{3}w_{3}&z_{3}&1  \end{array}\right)}
 \label{abcd}
 \end{eqnarray}
 
 In other words, there is a unique  map that connects two distinct set of triplets. Since any three points determine  a configuration of four mutually tangent circles), 
 any two Descartes configurations are related by a M\"{o}bius map. 

\section{ Proof of Descartes's Theorem for Ford Circles by Richard Friedberg\cite{RF} }

Theorem:  If three variables $\lambda_1$, $\lambda_2$ and $\lambda_3$  are related by,
\begin{equation}
\lambda_1 \pm \lambda_2 \pm \lambda_3 =0
\end{equation}
Then they also satisfy,
\begin{equation}
 ( \lambda1^2 + \lambda_2^2 + \lambda_3^2)^2 = 2 ( \lambda_1^4 + \lambda_2^4+ \lambda_3^4).
 \label {lam}
 \end{equation}
  Proof:  Define the standard three homogeneous symmetric polynomials on three variables,
  \begin{eqnarray}
 P_1  =  \lambda_1 + \lambda_2 + \lambda_3,\,\,\
 P_2  =  \lambda_1 \lambda_2 + \lambda_2 \lambda_3 + \lambda_3 \lambda_2 ,\,\,\ 
 P_3  =  \lambda_1 \lambda_2 \lambda_3,
 \label{s3}
 \end{eqnarray}
 and some additional ones that will be useful,
  \begin{eqnarray}
 T_2  =  \lambda_1^2 + \lambda_2^2 + \lambda_3^2 ,\,\,\
  T_4   =  \lambda_1^4 + \lambda_2^4 + \lambda_3^4 ,\,\,\
  U  =   \lambda_1^2 \lambda_2^2 + \lambda_2^2 \lambda_3^2 + \lambda_3^2 \lambda_2^2
 \label{U}
 \end{eqnarray}
 By direct multiplication, we have,
  \begin{eqnarray}
 P_1^2   =  T_2 + 2P_2 ,\,\,\
  T_2^2  =  T_4 + 2U ,\,\,\
 P_2^2   =  U + 2 P_1 P_3.
 \label{m3}
 \end{eqnarray}
  We now set $P_1=0$, that gives $T_2 = -2 P_2$. Squaring both sides, and then eliminating U we get,
  \begin{eqnarray}
 T_2 ^2  =  4 P_2^2 
   = 4 U
  = 2 T_2^2 -2T_4
 = 2 T_4
  \label{ed}
  \end{eqnarray}
  $T_2^2= 2T_4$ gives the required equation ( \ref{lam}).  Applying the above formulas for  the Farey relation $ q_c = q_R+q_L$,  we get the Descartes theorem for Ford circles.\\
  
  The above proof may be hinting at a much-desired elegant proof of Descartes' theorem, as a Ford Apollonian can be transformed into a general Apollonian through a M\"{o}bius transformation.  The details of this approach remain to be worked out. However,  in the special case where any Ford Apollonian is transformed into another Apollonian with one of the circles being $\kappa=1$ using the Equation (\ref{cmap}), the curvatures of the
 other three circles are $\kappa_x = ( p_x2+q_x^2-1)$, ( x= L,c,R). It is easy to show that these four circles satisfy Descartes' theorem—Equation (\ref{Q4})—provided the triplet $(\frac{p_L}{q_L}, \frac{p_c}{q_c}, \frac{p_R}{q_R})$ satisfies the Farey sum formula—Equation (\ref{FR}).

 \end{document}